\documentclass[aps,preprint]{revtex4}%
\usepackage{amsfonts}
\usepackage{amsmath}
\usepackage{amssymb}
\usepackage{graphicx}%
\begin{document}
\title{The degenerate Fermi gas with renormalized density-dependent interactions in
the K harmonic approximation}
\author{Seth T. Rittenhouse}
\affiliation{Department of Physics and JILA, University of Colorado, Boulder CO, 80309-0440}
\author{Chris H. Greene}
\affiliation{Department of Physics and JILA, University of Colorado, Boulder CO, 80309-0440}
\date{\today}
\pacs{03.75.Ss; 31.15.Ja}

\begin{abstract}
We present a simple implementation of a density-dependent, zero-range
interactions in a degenerate Fermi gas described in hyperspherical
coordinates. The method produces a 1D effective potential which accurately
describes the ground state energy as a function of the hyperradius, the rms
radius of the two spin component gas throughout the unitarity regime. In the
unitarity regime the breathing mode frequency is found to limit to the
non-interacting value. A dynamical instability, similar to the Bosenova, is
predicted to be possible in gases containing more than three spin components,
for large, negative, two-body scattering lengths.

\end{abstract}
\maketitle


\section{Introduction}

The use of zero-range contact interactions to model real interactions in
atomic systems has a long history \cite{fermi1936}. The interest in these
interaction models arises from the simplifications that can be made to complex
systems \cite{houbiers1997ssa,bruun1998ifg,roth2001esa}. Unfortunately, in
strongly interacting or high density systems, the overly singular nature of
the $\delta$-function often poses a problem. For example, when the
non-regularized zero-range interaction is used variationally in a two
component degenerate Fermi gas, it produces a collapse behavior that is not
seen in experiment
\cite{regal2004orc,kinast2004esr,zwierlein2004cpf,bartenstein2004cmb,bourdel2004esb}%
. One method of avoiding these problems is to use a renormalized interaction.
Ref. \cite{vonstecher2006renorm} does just that, introducing a
density-dependent interaction strength for a zero-range interaction. Our study
applies the hyperspherical K-harmonic method of Ref. \cite{rittenhouse2006hdd}
to this interaction.

The starting point of the K harmonic method describes the degenerate Fermi gas
with a set of $3N-1$ hyperangular coordinates on the surface of a $3N$
dimensional hypersphere of hyperradius $R$, where $N$ is the number of atoms
in the system. For this system, $R$ is simply the rms radius of the gas, but
more generally $R^{2}$ is proportional to the trace of the moment of inertia
tensor \cite{avery,SmirnovShitikova,fano1986aca}. This formulation is a
variational treatment of the $N$-body problem in which the hyperangular
behavior of the system is approximated by that of a non-interacting degenerate
Fermi gas. At first glance this approach might seem non-intuitive, but it is
natural to assume that, in a first approximation, the behavior of the gas will
be determined by its overall spatial extent. This type of approach has been
used in studying Bose-Einstein condensates \cite{bohn_esry_greene_hsbec} and
it has also been applied to finite nuclei \cite{SmirnovShitikova,Timofeyuk04}.
The theoretical approach developed here shares some mathematical kinship with
D-dimensional perturbation theory \cite{dimensionalPT}; for instance, the
$N\rightarrow\infty$ and $D\rightarrow\infty$ limits both result in
wavefunctions perfectly localized in the hyperradius. However, our goals and
motivations differ for the most part from those of Ref. \cite{dimensionalPT}.

The paper is organized as follows: Section II reviews the formulation that
leads to a hyperradial effective potential; Section III shows how to take the
hyperangular matrix element of an operator in the large $N$ limit; Section IV
applies this method to the two component gas with zero-range,
density-dependent interaction of Ref. \cite{vonstecher2006renorm}. Sections
IV(A) and IV(B) examine the resulting effective potential in systems with
positive and negative two-body scattering lengths $a$; Section IV(C) explores
the unitarity regime when $a\rightarrow\pm\infty$; Section IV(D) analyzes the
low energy radial excitation frequency. Section V expands the treatment of
Section IV to an arbitrary number of spin components and briefly examines the
resulting effective potentials. Finally, Section VI summarizes the results and
discusses future avenues of study.

\section{Hyperspherical coordinates}

The hyperspherical formulation starts with the K harmonics description given
in Ref. \cite{rittenhouse2006hdd}. We briefly restate this formulation in
order to make this article self contained. We begin by considering $N$
identical fermionic atoms of mass $m$ in a spherically symmetric oscillator
trap with oscillator frequency $\omega$ distributed equally in two spin
substates. The governing Hamiltonian is%
\begin{equation}
H=-\dfrac{\hbar^{2}}{2m}\sum_{i=1}^{N}\nabla_{i}^{2}+\dfrac{1}{2}m\omega
^{2}\sum_{i=1}^{N}r_{i}^{2}+\sum_{i>j}U_{int}\left(  \vec{r}_{ij}\right)
\label{Ham1}%
\end{equation}
where $\vec{r}_{i}$ is a trap centered vector describing the position of the
$i$th atom and $\vec{r}_{ij}=\vec{r}_{i}-\vec{r}_{j}$ is the separation vector
between atoms $i$ and $j$. Transforming into hyperspherical coordinates this
Hamiltonian becomes%
\begin{equation}
H=-\dfrac{\hbar^{2}}{2M}\left(  \dfrac{1}{R^{3N-1}}\dfrac{\partial}{\partial
R}R^{3N-1}\dfrac{\partial}{\partial R}-\dfrac{\mathbf{\Lambda}^{2}}{R^{2}%
}\right)  +\dfrac{1}{2}M\omega^{2}R^{2}+V_{int}\left(  R,\Omega\right)
\label{Ham2}%
\end{equation}
where $M=Nm$ and the generalized angular momentum operator $\mathbf{\Lambda}$
is defined by \cite{avery}%
\begin{align}
\mathbf{\Lambda}^{2}  &  =-\sum_{i>j}\Lambda_{ij}^{2},\label{hypermom}\\
\text{where }\Lambda_{ij}  &  =x_{i}\dfrac{\partial}{\partial x_{j}}%
-x_{j}\dfrac{\partial}{\partial x_{i}}.
\end{align}
The hyperradius $R$ is given as%
\begin{equation}
R\equiv\left(  \dfrac{1}{N}\sum_{i=1}^{N}r_{i}^{2}\right)  ^{1/2}.
\label{hyperr}%
\end{equation}
The remaining $3N-1$ degrees of freedom are defined by angular coordinates
$2N$ of which are the normal spherical polar angles for each atom $\left(
\phi_{1},\theta_{1},\phi_{2},\theta_{2},\ldots,\phi_{N},\theta_{N}\right)  $.
The remaining $N-1$ hyperangles are defined in the convention of
\cite{SmirnovShitikova} as%
\begin{align}
\tan\alpha_{i}  &  =\dfrac{\sqrt{\sum_{j=1}^{i}r_{j}^{2}}}{r_{i+1}%
},\label{hypera}\\
i  &  =1,2,3,...,N-1.\nonumber
\end{align}
Alternatively we may write this as%
\begin{align}
r_{n}  &  =\sqrt{N}R\cos\alpha_{n-1}\prod\limits_{j=n}^{N-1}\sin\alpha
_{j}\label{hypera2}\\
0  &  \leq\alpha_{j}\leq\dfrac{\pi}{2},\text{ }j=1,2,...,N-1\nonumber
\end{align}
where we define $\cos\alpha_{0}\equiv1$ and $\prod\limits_{j=N}^{N-1}%
\sin\alpha_{j}\equiv1$. Collectively the full set of hyperangles are referred
to as $\Omega$. For notational simplicity we have rewritten the interaction as
a function of the hyperradius and hyperangles.%
\begin{equation}
V_{int}\left(  R,\Omega\right)  =\sum_{i>j}U_{int}\left(  \vec{r}_{ij}\right)
\label{Vint}%
\end{equation}
We now make the assumption that the wave function for this system is
approximately separable into hyperradial and hyperangular parts,
\begin{equation}
\Psi=F\left(  R\right)  \Phi_{\lambda}\left(  \Omega,\sigma_{1},\sigma
_{2},\ldots,\sigma_{N}\right)  . \label{sep}%
\end{equation}
where $F\left(  R\right)  $ is an undetermined hyperradial function, $\left(
\sigma_{1},\sigma_{2},\ldots,\sigma_{N}\right)  $ are the spin coordinates for
the $N$ atoms and $\Phi_{\lambda}\left(  \Omega,\sigma_{1},\sigma_{2}%
,\ldots,\sigma_{N}\right)  $ is an eigenfunction of the hyperangular momentum
operator $\mathbf{\Lambda}^{2}$ which obeys the eigenvalue equation
$\mathbf{\Lambda}^{2}\Phi_{\lambda}\left(  \Omega\right)  =\lambda\left(
\lambda+3N-2\right)  \Phi_{\lambda}\left(  \Omega\right)  $ with $\lambda$ an
integer. We note that $R$ is completely symmetric under all permutations of
atomic space and spin coordinates, whereby all of the permutational symmetry
must be contained in $\Phi_{\lambda}$. Thus we will assume that $\Phi
_{\lambda}\left(  \Omega\right)  $ has the lowest value of $\lambda$ allowed
by the fermionic symmetry constraints. Eq. \ref{sep} can be viewed as a trial
wavefunction whose hyperradial behavior will be variationally optimized later.
To employ the variational principle we must consider the matrix element%
\[
\left\langle \Phi_{\lambda}\left\vert V_{int}\left(  R,\Omega\right)
\right\vert \Phi_{\lambda}\right\rangle
\]
where the integral is taken over all hyperangular coordinates. We now have
created an effective 1D Schr\"{o}dinger equation in the hyperradius%
\begin{equation}
\left(  \dfrac{-\hbar^{2}}{2M}\dfrac{\partial^{2}}{\partial R^{2}}%
+V_{eff}\left(  R\right)  \right)  R^{\left(  3N-1\right)  /2}F\left(
R\right)  =ER^{\left(  3N-1\right)  /2}F\left(  R\right)  \label{Ham3}%
\end{equation}
where the first derivative terms in Eq. \ref{Ham2} have been removed by
multiplying $F\left(  R\right)  $ by $R^{\left(  3N-1\right)  /2}$ and the
effective potential is given by%
\begin{equation}
V_{eff}\left(  R\right)  =\dfrac{\left(  3N-1\right)  \left(  3N-3\right)
}{8MR^{2}}+\dfrac{\lambda\left(  \lambda+3N-2\right)  }{2MR^{2}}+\dfrac{1}%
{2}M\omega^{2}R^{2}+\left\langle \Phi_{\lambda}\left\vert V_{int}\left(
R,\Omega\right)  \right\vert \Phi_{\lambda}\right\rangle . \label{Veff1}%
\end{equation}
In order to calculate $V_{eff}$ we must first specify the function
$\Phi_{\lambda}\left(  \Omega\right)  $. For $N$-body systems having
completely filled shells (magic numbers), to which we restrict this study,
this is given in Ref. \cite{rittenhouse2006hdd} as the Slater determinant:%
\begin{equation}
\Phi_{\lambda}\left(  \Omega,\sigma_{1},\sigma_{2},\ldots,\sigma_{N}\right)
=\dfrac{1}{\sqrt{N!}G\left(  R\right)  }\sum_{P}\left(  -1\right)
^{p}P\left[  \prod\limits_{j=1}^{N}\psi_{n_{j}\ell_{j}m_{j}}\left(  \vec
{r}_{j}\right)  \left\langle \sigma_{j}|m_{s_{j}}\right\rangle \right]  .
\label{phi}%
\end{equation}
Here the sum is over all possible permutations of the $N$ spatial and spin
coordinates. For brevity of notation the spin coordinates will be omitted,
i.e. we abbreviate $\Phi_{\lambda}\left(  \Omega\right)  =\Phi_{\lambda
}\left(  \Omega,\sigma_{1},\sigma_{2},\ldots,\sigma_{N}\right)  $. Here
$\psi_{n_{j}\ell_{j}m_{j}}\left(  \vec{r}_{j}\right)  $ is the spatial wave
function of the $j$th atom given by
\begin{equation}
r\psi_{n\ell m}\left(  \vec{r}\right)  =B_{\ell nl}\left(  \dfrac{r}%
{l}\right)  ^{\ell+1}L_{n}^{\ell+1/2}\left(  \dfrac{r^{2}}{l^{2}}\right)
\exp\left(  -r^{2}/2l^{2}\right)  Y_{\ell m}\left(  \omega\right)
\label{NIsol}%
\end{equation}
where $Y_{\ell m}\left(  \omega\right)  $ is a normal 3D spherical harmonic of
the solid angle $\omega$ and $l$ is the length scale of the oscillator
functions. $G\left(  R\right)  $ is the nodeless hyperradial solution to the
non-interacting $N$ particle Schr\"{o}dinger equation:
\begin{equation}
R^{\left(  3N-1\right)  /2}G\left(  R\right)  =A_{N}\exp\left(  -R^{2}%
/2\mathcal{L}^{2}\right)  \left(  R/\mathcal{L}\right)  ^{\lambda+3N/2-1/2}
\label{oscRfun}%
\end{equation}
with $\mathcal{L}=l/\sqrt{N}$ and the spin ket $\left\vert m_{s}\right\rangle
$ allows for different spin components. While $\Phi_{\lambda}$ is constructed
from independent-particle oscillator functions, it is completely independent
of the length scale given by $l$. To simplify the overall behavior we will
only consider filled energy shells, the so called magic numbers, of atoms. We
will also be particularly interested in the large $N$ limit of the system.
Finally, to simplify the procedure we rescale $R$ and $E$ in Eq. \ref{Ham3}
their values $R_{NI}\equiv\sqrt{\left\langle R^{2}\right\rangle _{NI}}$ and
$E_{NI}$ for the noninteracting $N$-particle oscillator:%
\begin{align}
E  &  =E_{NI}E^{\prime}\label{ERrescale}\\
R  &  =\sqrt{\left\langle R^{2}\right\rangle _{NI}}R^{\prime},\nonumber
\end{align}
which introduces the dimensionless variables of energy ($E^{\prime}$) and
hyperradius ($R^{\prime}$). Here the non-interacting energy $E_{NI}$ and
average hyperradius squared $\left\langle R^{2}\right\rangle _{NI}$ are given
explicitly by%
\begin{align*}
E_{NI}  &  =\left(  \lambda+\dfrac{3N}{2}\right)  \hbar\omega\\
\left\langle R^{2}\right\rangle _{NI}  &  =\left(  \dfrac{\lambda}{N}%
+\dfrac{3}{2}\right)  l_{0}^{2}.
\end{align*}
Here $l_{0}=\sqrt{\hbar/m\omega}$ is the one particle oscillator length. Under
this rescaling the effective Schrodinger equation becomes%
\begin{equation}
\left(  \dfrac{-1}{2m^{\ast}}\dfrac{\partial^{2}}{\partial R^{\prime2}}%
+\dfrac{V_{eff}\left(  R^{\prime}\right)  }{E_{NI}}\right)  R^{\prime\left(
3N-1\right)  /2}F\left(  R^{\prime}\right)  =E^{\prime}R^{\prime\left(
3N-1\right)  /2}F\left(  R^{\prime}\right)  \label{Hrescal}%
\end{equation}
with $m^{\ast}=mE_{NI}N\left\langle R^{2}\right\rangle _{NI}/\hbar^{2}$. Exact
values of $\lambda$ and $N$ for the $n$th filled shell are written in Refs.
\cite{roth2001esa,rittenhouse2006hdd} , the large $N$ limit of interest here
is approximated by%
\[
\lambda\rightarrow\dfrac{\left(  3N\right)  ^{4/3}}{4}%
\]
giving the effective potential as%
\begin{equation}
\dfrac{V_{eff}\left(  R^{\prime}\right)  }{E_{NI}}\rightarrow\dfrac
{1}{2R^{\prime2}}+\dfrac{1}{2}R^{\prime2}+\dfrac{\left\langle \Phi_{\lambda
}\left\vert V_{int}\left(  R^{\prime},\Omega\right)  \right\vert \Phi
_{\lambda}\right\rangle }{E_{NI}}. \label{Veffrescale}%
\end{equation}
Here $V_{int}\left(  R^{\prime},\Omega\right)  $ is the interaction potential
of Eq. \ref{Vint} written in terms of the rescaled hyperradius. We now need a
method for calculating the interaction matrix element in the large $N$ limit.

\section{Operator matrix elements in the $N\rightarrow\infty$ limit}

In this section we develop a method for calculating hyperangular matrix
elements of an operator in the large $N$ limit, e.g.,
\begin{equation}
\tilde{O}\left(  R^{\prime}\right)  =\int\Phi_{\lambda}\left(  \Omega\right)
O\left(  R^{\prime},\Omega\right)  \Phi_{\lambda}\left(  \Omega\right)
d\Omega. \label{Otil}%
\end{equation}
Here $O\left(  R^{\prime},\Omega\right)  $ is a general operator that is a
function of the rescaled hyperradius and the hyperangles. To allow us to
integrate over all of the $3N$ dimensions of the space, we multiply both sides
of Eq. \ref{Otil} by a $\delta$-function in the hyperradius and integrate.%
\begin{equation}
\tilde{O}\left(  R_{0}^{\prime}\right)  =\int\delta\left(  R^{\prime}%
-R_{0}^{\prime}\right)  \Phi_{\lambda}\left(  \Omega\right)  O\left(
R^{\prime},\Omega\right)  \Phi_{\lambda}\left(  \Omega\right)  d\Omega
dR^{\prime}. \label{OtilR0}%
\end{equation}
To create the $\delta$-function we consider a function of the form
\begin{equation}
R^{\prime\left(  3N-1\right)  /2}G_{N}\left(  R^{\prime}\right)  =A_{N}%
\exp\left(  -\dfrac{R^{\prime2}N\left\langle R^{2}\right\rangle _{NI}}%
{2l_{0}^{2}R_{0}^{\prime2}}\right)  \left(  \dfrac{\sqrt{N\left\langle
R^{2}\right\rangle _{NI}}R^{\prime}}{l_{0}R_{0}^{\prime}}\right)
^{\lambda+3N/2-1/2} \label{limfun}%
\end{equation}
where $A_{N}$ is a normalization constant and $l_{0}=\sqrt{\hbar/m\omega}$ is
the oscillator length. In the limit where $N\rightarrow\infty$ we see that
using this definition we have that%
\[
\lim\limits_{N\rightarrow\infty}\left[  R^{\prime\left(  3N-1\right)  /2}%
G_{N}\left(  R^{\prime}\right)  \right]  ^{2}=\delta\left(  R^{\prime}%
-R_{0}^{\prime}\right)  .
\]
From this we make the substitution in Eq. \ref{OtilR0}
\begin{equation}
\tilde{O}\left(  R_{0}^{\prime}\right)  =\lim\limits_{N\rightarrow\infty}%
\int\left[  R^{\prime\left(  3N-1\right)  }G_{N}\left(  R^{\prime}\right)
\right]  ^{2}\Phi_{\lambda}\left(  \Omega\right)  O\left(  R^{\prime}%
,\Omega\right)  \Phi_{\lambda}\left(  \Omega\right)  d\Omega dR^{\prime}.
\label{OtilR02}%
\end{equation}
Referring to Eq. \ref{phi} and remembering that the K-harmonic $\Phi_{\lambda
}\left(  \Omega\right)  $ is \emph{independent} of the oscillator length
scale, it follows that the wave function $G_{N}\left(  R^{\prime}\right)
\Phi_{\lambda}\left(  \Omega\right)  $ is merely a Slater determinant of
non-interacting single particle oscillator states with oscillator length
\begin{equation}
l_{eff}=R_{0}^{\prime}l_{0}. \label{leff}%
\end{equation}
Further, Ref. \cite{avery} gives that $R^{\prime\left(  3N-1\right)
}dR^{\prime}d\Omega$ is the full volume element for the $3N$ dimensional
space. All of this implies that in the large $N$ limit, the hyperangular
operator expectation value $\left\langle \Phi_{\lambda}\left\vert O\left(
R,\Omega\right)  \right\vert \Phi_{\lambda}\right\rangle $ is approximated by
the full expectation value of the operator for a trial wavefunction consisting
of a Slater determinant of non-interacting oscillator states, i.e.%
\begin{equation}
\tilde{O}\left(  R_{0}^{\prime}\right)  =\left\langle D_{l_{eff}}\left(
\vec{r}_{1},\vec{r}_{2},...,\vec{r}_{N}\right)  \left\vert O\left(  R^{\prime
},\Omega\right)  \right\vert D_{l_{eff}}\left(  \vec{r}_{1},\vec{r}%
_{2},...,\vec{r}_{N}\right)  \right\rangle _{3N} \label{DetME}%
\end{equation}
where $D_{l_{eff}}\left(  \vec{r}_{1},\vec{r}_{2},...,\vec{r}_{N}\right)  $ is
a Slater determinant of oscillator states with oscillator length $l_{eff}$ and
the subscript $3N$ is to indicate that the matrix element is taken over all
$3N$ spatial and $N$ spin degrees of freedom.

\section{Renormalized zero-range interactions}

To show the utility of the result in the previous section, we will apply it to
the density-dependent renormalized zero-range interactions presented in Ref.
\cite{vonstecher2006renorm}, in which a zero-range interaction is used whose
strength is dependent on the density of the gas.
\begin{equation}
U_{int}\left(  \vec{r}_{ij}\right)  =\dfrac{4\pi\hbar^{2}}{m}\dfrac
{\zeta\left(  k_{f}\left(  \vec{r}_{i}\right)  a\right)  }{k_{f}\left(
\vec{r}_{i}\right)  }\delta\left(  \vec{r}_{ij}\right)  \label{renormint}%
\end{equation}
where $a$ is the two-body s-wave scattering length and the fermi wave number
$k_{f}=k_{f}\left(  \vec{r}\right)  =\left(  6\pi^{2}\rho^{\left(  1\right)
}\left(  \vec{r}\right)  \right)  ^{1/3}$ is defined in terms of the single
spin component density, $\rho^{\left(  1\right)  }\left(  \vec{r}\right)  $.
We approximate the dimensionless renormalized function $\zeta\left(
k_{f}a\right)  $ from Ref. \cite{vonstecher2006renorm} with%
\begin{align}
\zeta\left(  k_{f}a\right)   &  =A+B\arctan\left(  Ck_{f}a-D\right)
\label{gkfa}\\
&  \text{where}\nonumber\\
A  &  =0.3949\nonumber\\
B  &  =1.1375\nonumber\\
C  &  =\dfrac{1+\tan^{2}\left(  \dfrac{A}{B}\right)  }{B}=0.9942\nonumber\\
D  &  =\tan\left(  \dfrac{A}{B}\right)  =0.3618.\nonumber
\end{align}
Two of the fitting parameters $A$ and $B$ are found by fitting the asymptotic
behavior of $\zeta\left(  k_{f}a\right)  $ as $k_{f}a\rightarrow\pm\infty$,
which are given in ref. \cite{vonstecher2006renorm} by%
\begin{align*}
\lim\limits_{k_{f}a\rightarrow\infty}\zeta\left(  k_{f}a\right)   &  =2.1817\\
\lim\limits_{k_{f}a\rightarrow-\infty}\zeta\left(  k_{f}a\right)   &  =-1.3919
\end{align*}
The constants $C$ and $D$ in Eq. \ref{gkfa} are determined by matching the
Fermi pseudo-potential in the $\left\vert k_{f}a\right\vert \ll1$ limit
\cite{fermi1936,roth2001esa}, i.e.
\begin{equation}
\dfrac{4\pi\hbar^{2}}{m}\dfrac{\zeta\left(  k_{f}a\right)  }{k_{f}}%
\rightarrow\dfrac{4\pi\hbar^{2}a}{m}. \label{Fermipseudo}%
\end{equation}
\begin{figure}[ptb]
\begin{center}
\includegraphics[width=3in]{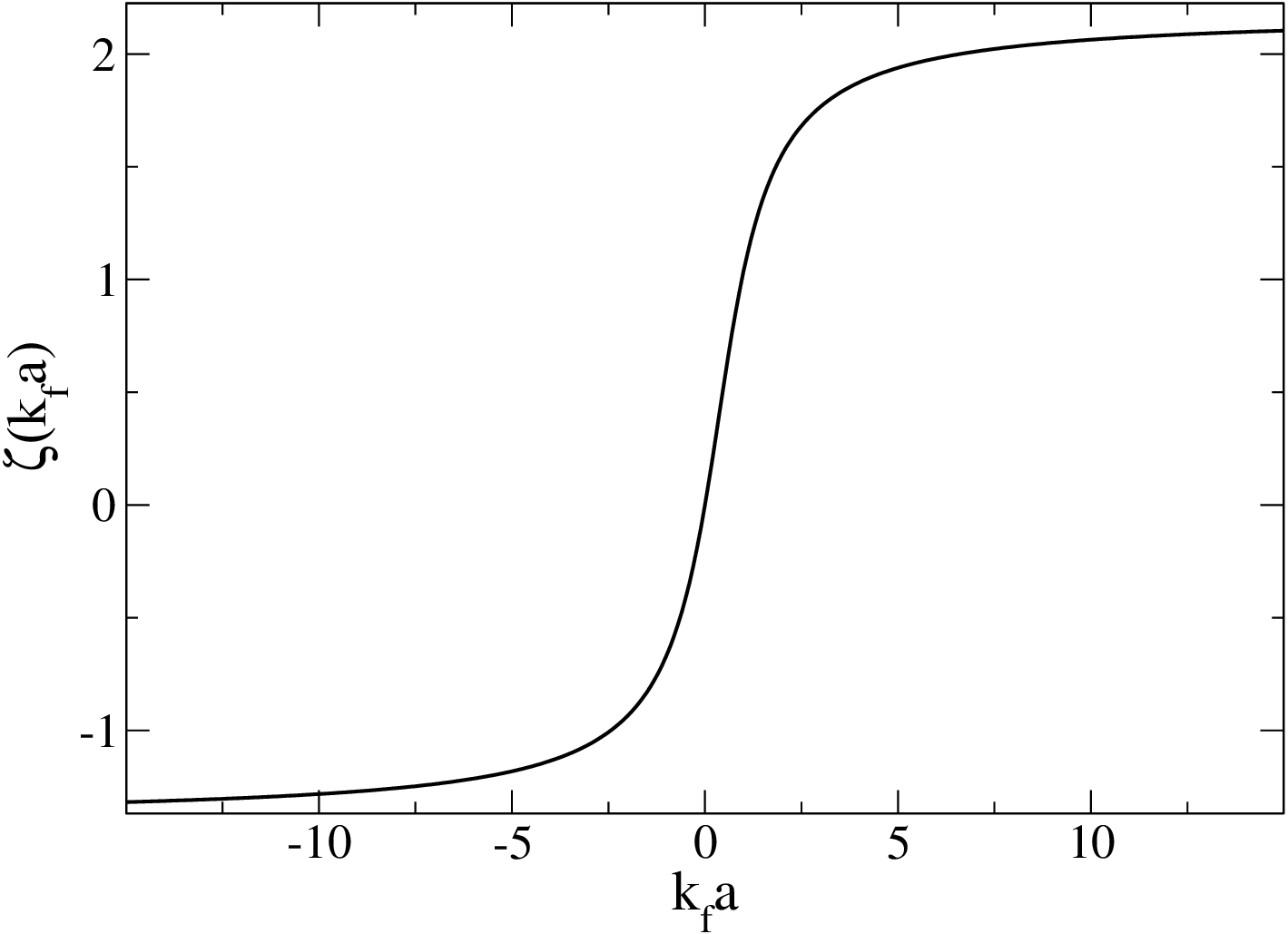}
\end{center}
\caption{The density-dependent interaction strength function $\zeta(k_{f}a)$
is shown plotted versus $k_{f}a$.}%
\label{zetaeff}%
\end{figure}Fig. \ref{zetaeff} shows the behavior of this interaction as a
function of $k_{f}a$.

Eq. \ref{DetME} implies that%
\[
\left\langle \Phi_{\lambda}\left\vert V_{int}\left(  R^{\prime},\Omega\right)
\right\vert \Phi_{\lambda}\right\rangle =\left\langle D_{l_{efff}}\left(
\vec{r}_{1},\vec{r}_{2},...,\vec{r}_{N}\right)  \left\vert \sum_{i>j}%
U_{int}\left(  \vec{r}_{ij}\right)  \right\vert D_{l_{eff}}\left(  \vec{r}%
_{1},\vec{r}_{2},...,\vec{r}_{N}\right)  \right\rangle _{3N}%
\]
Owing to the exchange anti-symmetry of the determinantal wavefunction and the
orthogonality of the single atom wave functions (see Ref. \cite{cowan1981tas}
for details), this becomes%
\begin{align*}
\left\langle \Phi_{\lambda}\left\vert V_{int}\left(  R^{\prime},\Omega\right)
\right\vert \Phi_{\lambda}\right\rangle  &  =\dfrac{1}{2}\sum_{i,j=1}%
^{N}\left[  \int\left\vert \psi_{i}\left(  \vec{r}_{1}\right)  \right\vert
^{2}U_{int}\left(  \vec{r}_{12}\right)  \left\vert \psi_{j}\left(  \vec{r}%
_{2}\right)  \right\vert ^{2}d^{2}r_{1}d^{3}r_{2}\right. \\
&  -\left.  \delta_{m_{s_{i}}m_{s_{j}}}\int\psi_{i}^{\ast}\left(  \vec{r}%
_{1}\right)  \psi_{j}\left(  \vec{r}_{1}\right)  U_{int}\left(  \vec{r}%
_{12}\right)  \psi_{j}^{\ast}\left(  \vec{r}_{2}\right)  \psi_{i}\left(
\vec{r}_{2}\right)  d^{3}r_{1}d^{3}r_{2}\right]
\end{align*}
Here $\psi_{i}\left(  \vec{r}\right)  $ is the spatial state of an atom in the
$i$th orbital in the spin substate defined by $m_{s_{i}}$ and the sum runs
over all the single particle states in the original determinant. Substituting
Eq. \ref{renormint} for the interaction and using the $\delta$-function to
simplify one of the integrals gives%
\begin{equation}
\left\langle \Phi_{\lambda}\left\vert V_{int}\left(  R^{\prime},\Omega\right)
\right\vert \Phi_{\lambda}\right\rangle =\dfrac{1}{2}\dfrac{4\pi\hbar^{2}}%
{m}\sum_{i,j=1}^{N}\left(  1-\delta_{m_{s_{i}}m_{s_{j}}}\right)
\int\left\vert \psi_{i}\left(  \vec{r}_{1}\right)  \right\vert ^{2}\left\vert
\psi_{j}\left(  \vec{r}_{1}\right)  \right\vert ^{2}\dfrac{\zeta\left(
k_{f}\left(  \vec{r}_{1}\right)  a\right)  }{k_{f}\left(  \vec{r}_{1}\right)
}d^{3}r_{1} \label{Vint1}%
\end{equation}
If we sum over all possible spin projections $\delta_{m_{s_{i}}m_{s_{j}}}$ and
remember that we have assumed an equal distribution of atoms in each spin
substate we arrive at%
\begin{equation}
\left\langle \Phi_{\lambda}\left\vert V_{int}\left(  R^{\prime},\Omega\right)
\right\vert \Phi_{\lambda}\right\rangle =\dfrac{4\pi\hbar^{2}}{m}\int
\dfrac{\zeta\left[  k_{f}\left(  \vec{r}\right)  a\right]  }{k_{f}\left(
\vec{r}\right)  }\left[  \rho_{l_{eff}}^{\left(  1\right)  }\left(  \vec
{r}\right)  \right]  ^{2}d^{3}r. \label{Intelem}%
\end{equation}
Here we have used the definition of the density of a single spin component
\[
\rho_{l_{eff}}^{\left(  1\right)  }\left(  \vec{r}\right)  =\sum_{i}%
^{N/2}\left\vert \psi_{n_{i}\ell_{i}m_{i}}\left(  \vec{r}\right)  \right\vert
^{2}.
\]
In the Thomas-Fermi approximation, which should be exact for non-interacting
oscillator states in the large $N$ limit, this density is given in oscillator
units ($l_{0}=\hbar\omega=1$) by%
\begin{equation}
\rho_{l_{eff}}^{\left(  1\right)  }\left(  \vec{r}\right)  =\dfrac{1}{6\pi
^{2}l_{eff}^{3}}\left(  2\mu\right)  ^{3/2}\left(  1-\dfrac{r^{2}}%
{2l_{eff}^{2}\mu}\right)  ^{3/2} \label{dense}%
\end{equation}
where $\mu=\left(  3N\right)  ^{1/3}$ is the chemical potential at zero
temperature of $N$ non-interacting fermions divided equally between two
different spin substates. We may also note from Eq. \ref{leff} that in
oscillator units, $l_{eff}=R^{\prime}.$ Inserting Eq. \ref{gkfa} and
$k_{f}\left(  \vec{r}\right)  =\left[  6\pi^{2}\rho_{leff}^{\left(  1\right)
}\left(  \vec{r}\right)  \right]  ^{1/3}$ and making a change of variables in
the integral, Eq. \ref{Intelem} becomes%
\begin{align}
\left\langle \Phi_{\lambda}\left\vert V_{int}\left(  R_{0}^{\prime}%
,\Omega\right)  \right\vert \Phi_{\lambda}\right\rangle  &  =\dfrac{64N^{4/3}%
}{3^{2/3}\pi^{2}R^{\prime2}}f\left(  \dfrac{k_{f}^{0}a}{R^{\prime}}\right)
,\label{Vintfinal}\\
\text{where }f\left(  \dfrac{k_{f}^{0}a}{R^{\prime}}\right)   &  \equiv
\int_{0}^{1}y^{6}\sqrt{1-y^{2}}\zeta\left(  \dfrac{k_{f}^{0}a}{R^{\prime}%
}y\right)  dy.\nonumber
\end{align}
Here $k_{f}^{0}$ is the peak Fermi wave number for $N$ non-interacting atoms,
i.e. $k_{f}^{0}=\left[  6\pi^{2}\rho_{l_{0}}^{\left(  1\right)  }\left(
0\right)  \right]  ^{1/3}=\sqrt{2\left(  3N\right)  ^{1/3}}$. Observe that the
only parameter in this expression is $k_{f}^{0}a$ which is dimensionless.
Inserting Eq. \ref{Vintfinal} into Eq. \ref{Veffrescale} now gives the final
effective hyperradial potential in the $N\gg1$ limit.%
\begin{equation}
\dfrac{V_{eff}\left(  R^{\prime}\right)  }{E_{NI}}\rightarrow\dfrac
{1}{2R^{\prime2}}+\dfrac{1}{2}R^{\prime2}+\dfrac{256}{9\pi^{2}R^{\prime2}%
}f\left(  \dfrac{k_{f}^{0}a}{R^{\prime}}\right)  . \label{Vefffinal}%
\end{equation}

For $\left\vert k_{f}^{0}a\right\vert \ll R^{\prime}$ the integral may be
evaluated exactly using Eq. \ref{Fermipseudo} giving
\begin{equation}
\dfrac{V_{eff}\left(  R^{\prime}\right)  }{E_{NI}}\rightarrow\dfrac
{1}{2R^{\prime2}}+\dfrac{1}{2}R^{\prime2}+\dfrac{4096k_{f}^{0}a}{2835\pi
^{2}R^{\prime3}} \label{Veffbare}%
\end{equation}
which is exactly the behavior predicted in Ref. \cite{rittenhouse2006hdd}
using non-renormalized zero-range interactions.

\subsection{Repulsive effective interactions, $a>0$}

\begin{figure}[ptb]
\begin{center}
\includegraphics[width=3in]{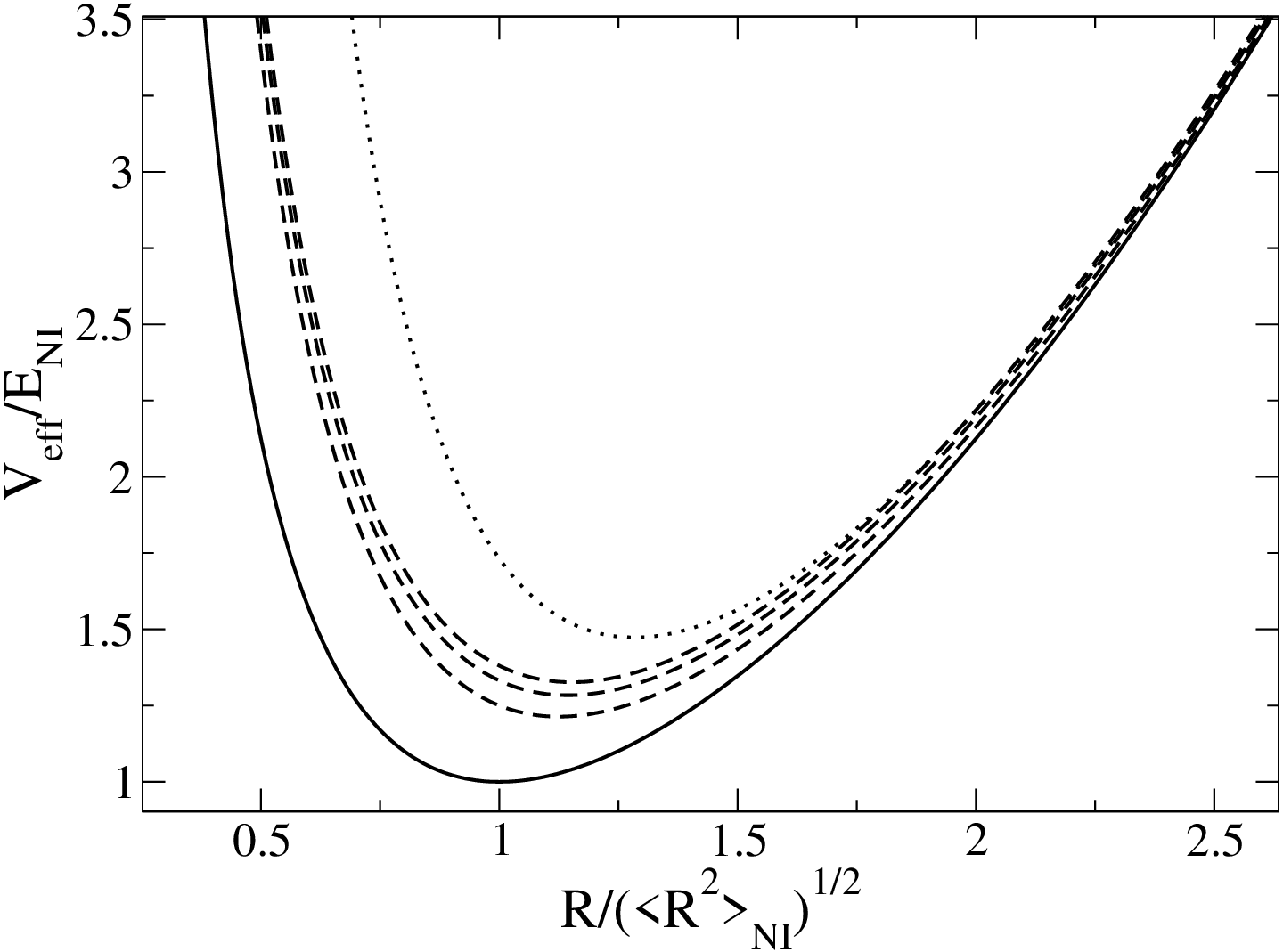}
\end{center}
\caption{The dimensionless ratio of the effective hyperradial potential to the
non-interacting total energy is plotted as a function of the dimensionless
rescaled hyperradius, for several different repulsive interaction strengths.
The non-interacting limit $k_{f}^{0}a=0$ is shown as the sold curve; the
dashed curves show the renormalized effective potential for (bottom to top)
$k_{f}^{0}a=2$, $k_{f}^{0}a=5$ and $k_{f}^{0}a=50$. Also shown is the
non-renormalized effective potential with $k_{f}^{0}a=5$ (dotted curve).}%
\label{Veffrepul}%
\end{figure}

Here we explore the behavior of the DFG under a repulsive effective potential
where the two-body scattering length, $a$, is positive. The renormalized
description of the interactions used here and in Ref.
\cite{vonstecher2006renorm} is only accurate if the real two-body interactions
are purely repulsive or if the gas is somehow prevented from forming into
molecular dimer states. In other words we can only look at a gas of atoms not
of molecules. Fig. \ref{Veffrepul}, which shows $V_{eff}$ for several positive
two-body scattering lengths, also shows an example of the bare
non-renormalized effective potential. As one would expect, the repulsive
interactions cause the gas to push out against itself and against the trap
walls, which increases the overall energy and size of the gas.
\begin{figure}[ptbptb]
\begin{center}
\includegraphics[width=3in]{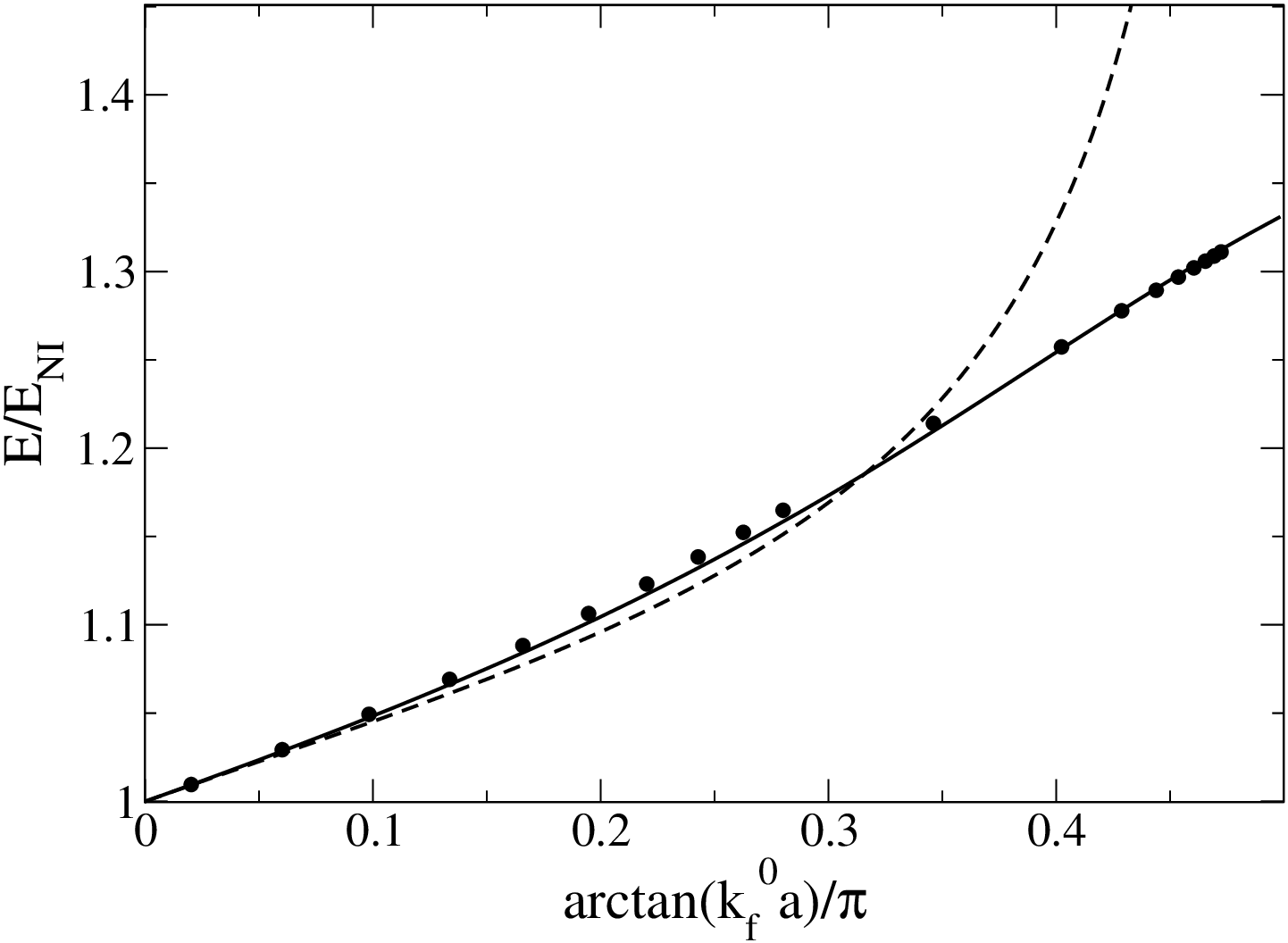}
\end{center}
\caption{The ground state energy of the DFG in units of the non-interacting
energy predicted by the K harmonic method (solid line) is plotted versus
$\arctan\left(  k_{f}^{0}a\right)  /\pi$ and compared with that predicted by
the Hartree-Fock method with 2280 atoms (circles). The dashed line is the
ground state energy predicted by the K harmonic method using the bare Fermi
pseudo-potential.}%
\label{Erepuls}%
\end{figure}\begin{figure}[ptbptbptb]
\begin{center}
\includegraphics[width=3in]{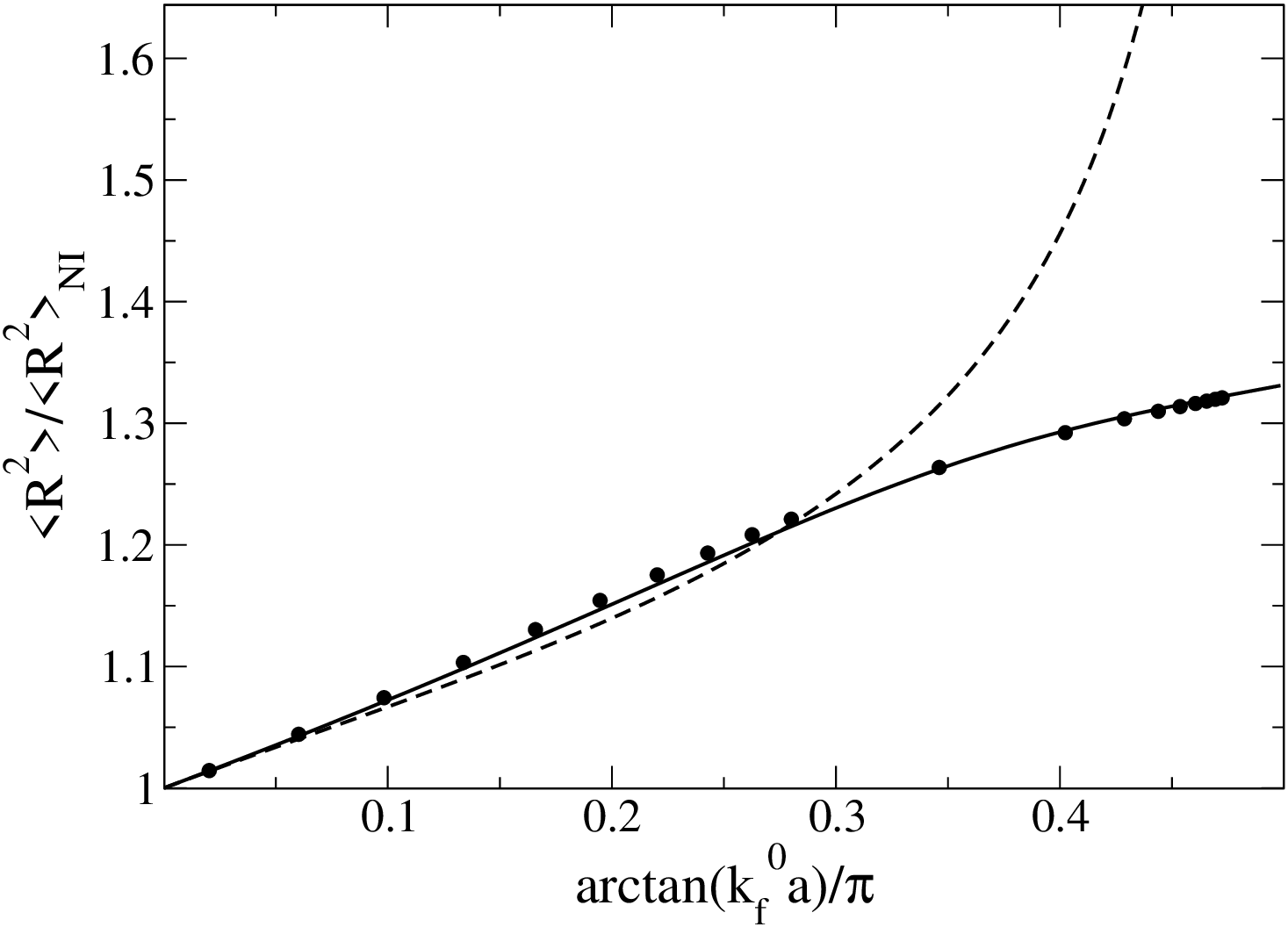}
\end{center}
\caption{The ground state average squared hyperradius of the two-component DFG
in the large-N limit, divided by the non-interacting value for this quantity,
is plotted versus $\arctan\left(  k_{f}^{0}a\right)  /\pi$. Also shown are the
values predicted by the Hartree-Fock method with 2280 atoms (circles). The
dashed line is the ground state energy predicted by the K harmonic method
using the bare Fermi pseudo-potential.}%
\label{Rrepuls}%
\end{figure}

The true ground state energy would be found by solving Eq. \ref{Hrescal} for
the lowest eigenvalue, but if we examine $m^{\ast}$ in the large $N$ limit one
can see that the second derivative term becomes negligible. The ground state
energy and hyperradius can thus be found by minimizing the effective potential
$V_{eff}\left(  R^{\prime}\right)  .$ Figs. \ref{Erepuls} and \ref{Rrepuls}
show the energy and average squared hyperradius of the minimum of $V_{eff}$ as
functions of $k_{f}^{0}a$, compared to those same values calculated using the
bare non-renormalized effective potential given by Eq. \ref{Veffbare}. Also
shown are the ground state energy and average hyperradius squared predictions
from the Hartree-Fock method using the renormalized interaction. As the
interaction gets stronger the renormalized energies and hyperradii flatten out
and approach a constant in the unitarity limit. This behavior will be examined
more carefully in a Section IV(C). For $k_{f}^{0}a\ll1$ the Fermi
pseudo-potential approximation is in good quantitative agreement with the
renormalized interactions, but diverges dramatically as $k_{f}^{0}%
a\rightarrow\infty$. This dramatizes the breakdown of the non-renormalized
zero-range approximation, which overestimates the interaction strength as the
unitarity regime is approached.

\subsection{Attractive effective interactions, $a<0$}

\begin{figure}[ptb]
\begin{center}
\includegraphics[width=3in]{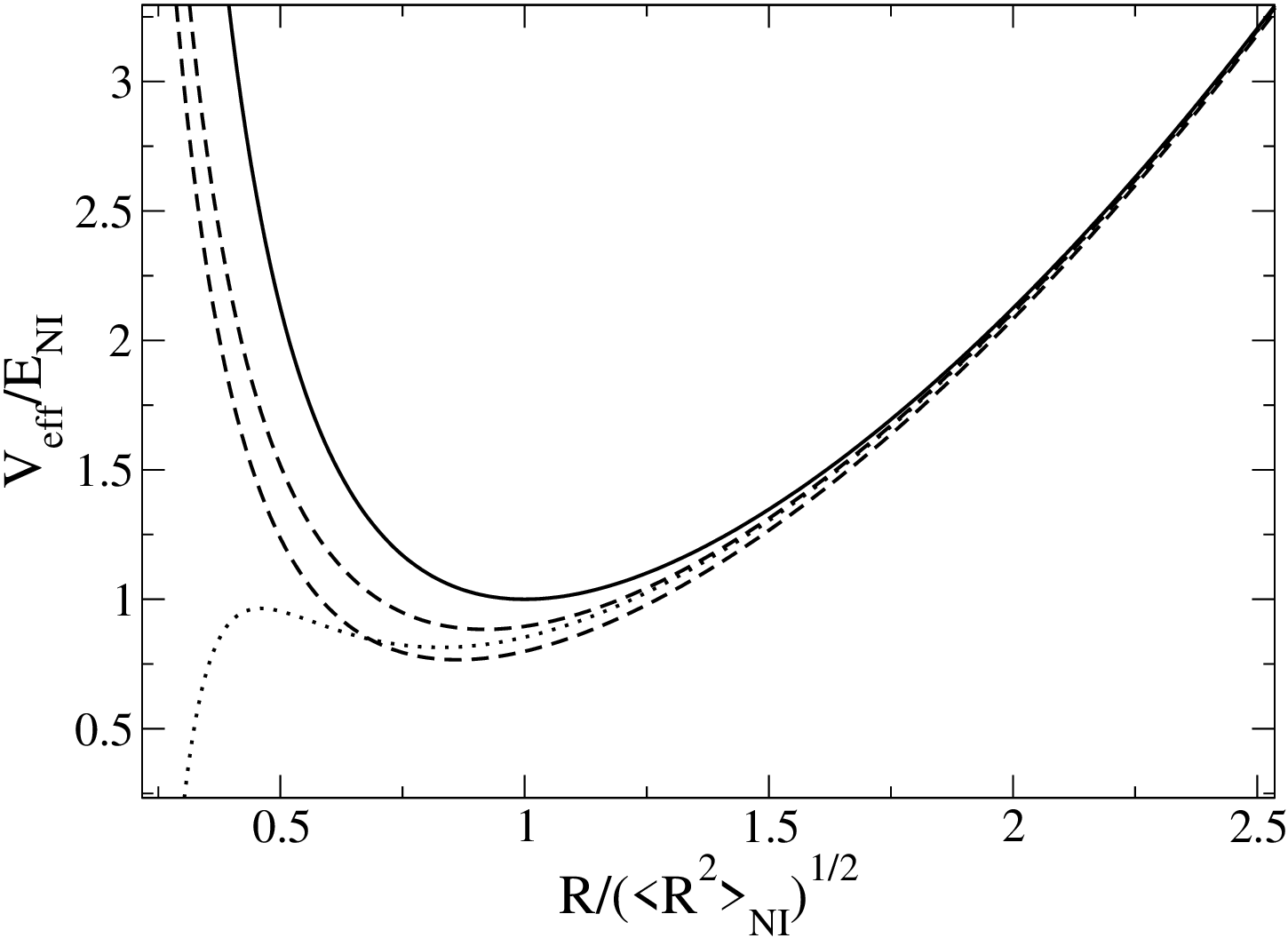}
\end{center}
\caption{The effective potential in units of the non-interacting energy is
plotted versus the hyperradius in units of $\sqrt{\left\langle R^{2}%
\right\rangle _{NI}}$, for several interaction strengths. The non-interacting
limit $k_{f}^{0}a=0$ is shown as the sold curve and the dashed curves show the
renormalized effective potential for (top to bottom) $k_{f}^{0}a=-1$ and
$k_{f}^{0}a=-5$. Also shown is the non-renormalized effective potential with
$k_{f}^{0}a=-1$ (dotted curve).}%
\label{Veffattract}%
\end{figure}

Fig. \ref{Veffattract} shows the behavior of the effective potential for some
attractive values of the two-body scattering length, along with an example of
the non-renormalized effective potential. The decisive qualitative importance
of the renormalization is now apparent; for attractive non-renormalized
interactions the interaction term in the effective potential will always take
over as $R^{\prime}\rightarrow0$ which creates an inner collapse region where
the ground state energy of the gas diverges toward $-\infty$ and the gas lives
in a metastable outer potential well. In contrast, the renormalized effective
potential has no such collapse phenomenon, and the ground state of the gas is
in a global minimum. This behavior will be discussed further in Section IV(C).

\begin{figure}[ptbptb]
\begin{center}
\includegraphics[width=3in]{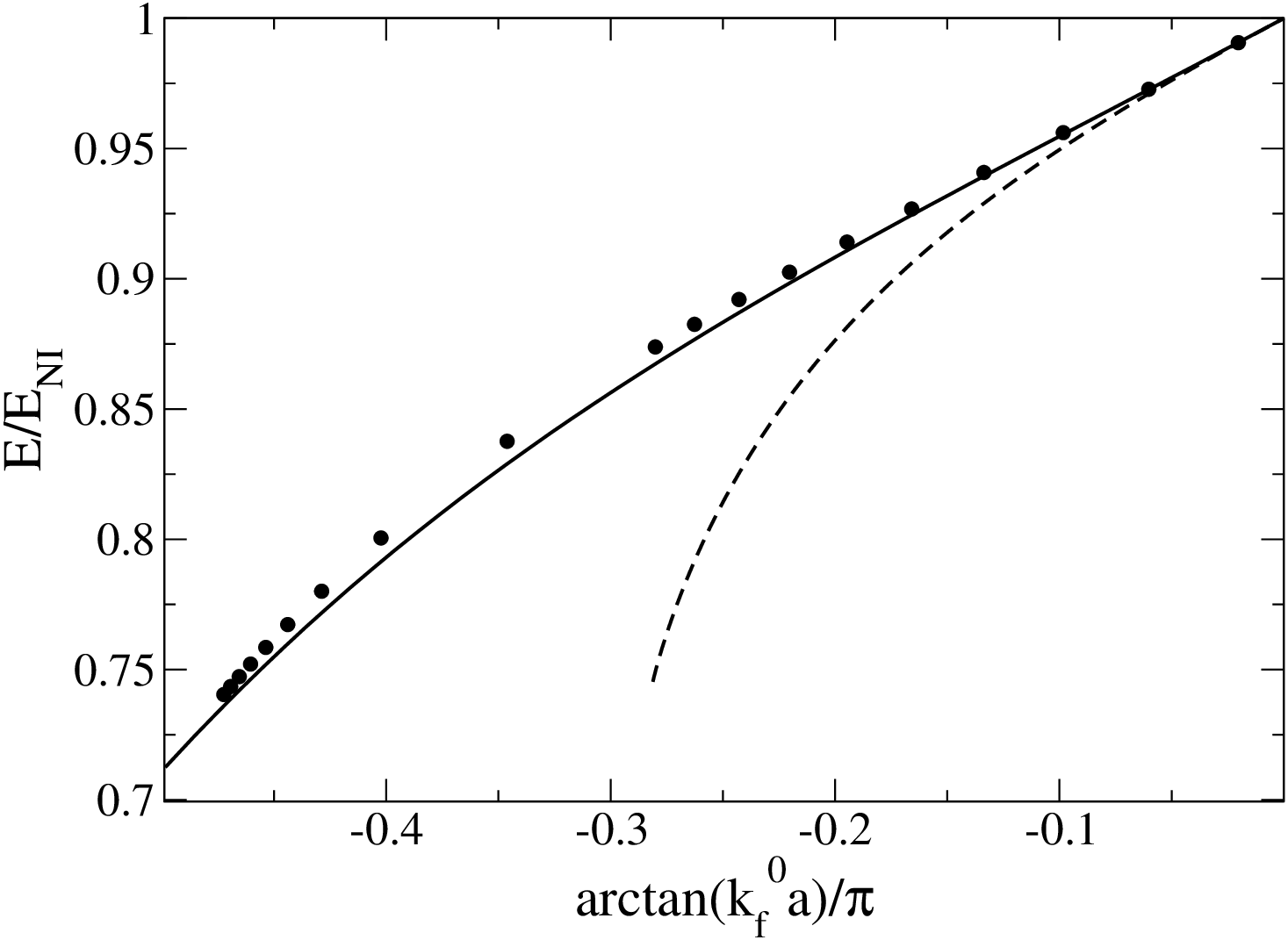}
\end{center}
\caption{The ground state energy of the DFG, predicted by the K harmonic
method in units of the non-interacting energy (solid line) is shown as a
function of $\arctan\left(  k_{f}^{0}a\right)  $/$\pi$ as a solid line. Also
shown is that predicted by the Hartree-Fock method for 2280 atoms (circles).
The dashed line is the ground state energy predicted by the K harmonic method
using the bare Fermi pseudopotential.}%
\label{Eattract}%
\end{figure}

\begin{figure}[ptbptbptb]
\begin{center}
\includegraphics[width=3in]{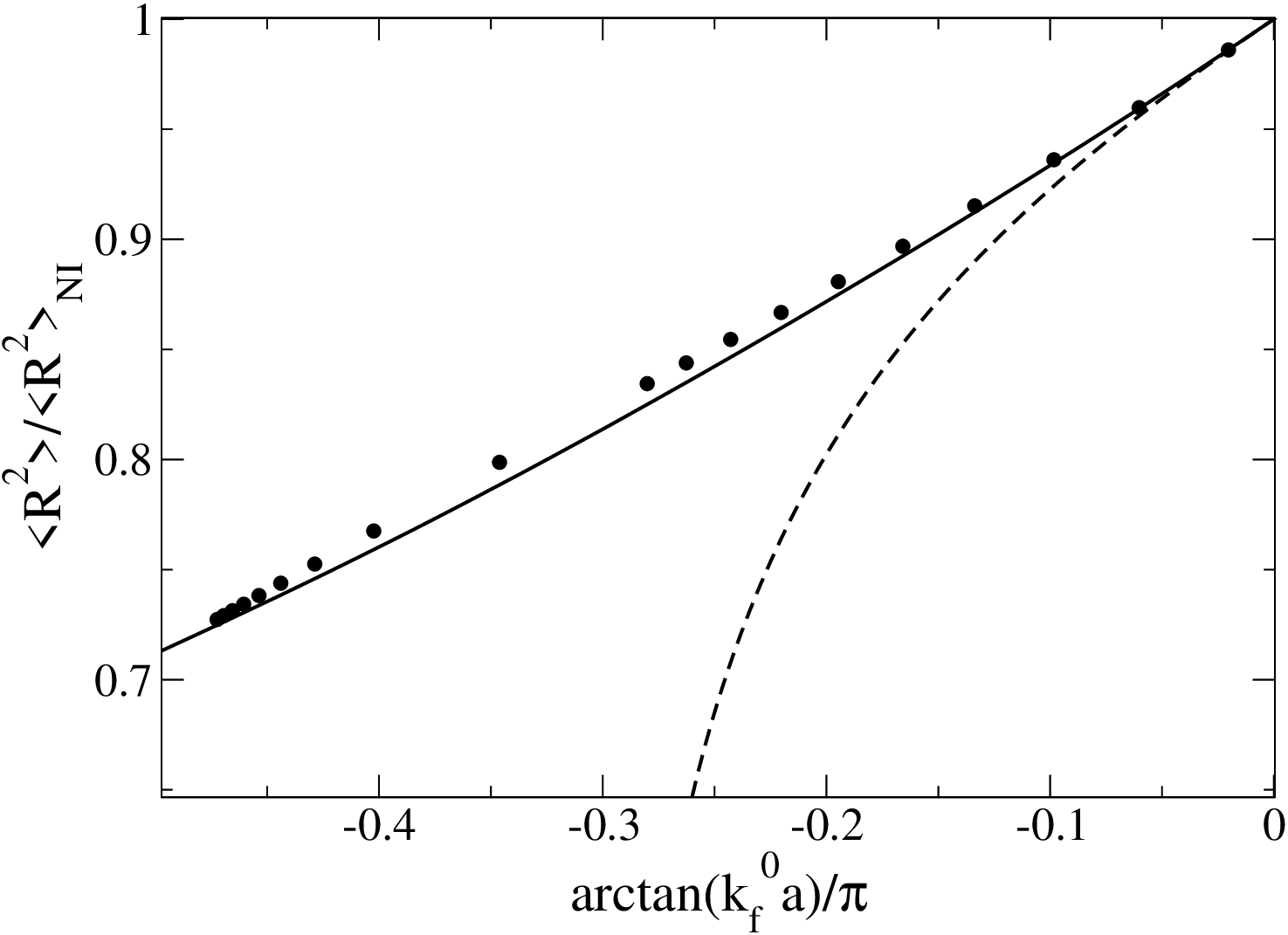}
\end{center}
\caption{The ground state average hyperradius squared of the DFG in units of
the non-interacting average hyperradius squared predicted by the K harmonic
method (solid line) is shown plotted against $\arctan\left(  k_{f}%
^{0}a\right)  /\pi$ compared with that predicted by the Hartree-Fock method
with 2280 atoms (circles). The dashed line is the ground state energy
predicted by the K harmonic method using the bare Fermi pseudopotential.}%
\label{Rattract}%
\end{figure}

Figs. \ref{Eattract} and \ref{Rattract} show the ground state energy and
average hyperradius squared of the system compared to the non-renormalized
values. The effects of renormalization for attractive interaction are even
more striking than in the repulsive interaction case. Without renormalization
the metastable region of the effective potential disappears for $k_{f}%
^{0}a<-1.21$ \cite{rittenhouse2006hdd} and the gas has no barrier to prevent
it from falling into the central collapse region. With renormalization, as
$k_{f}^{0}a\rightarrow-\infty$ the energy and average hyperradius squared go
towards a fixed value. Figs. \ref{Eattract} and \ref{Rattract} also show the
ground state energy and average squared hyperradius predictions from the
Hartree-Fock method. For $\left\vert k_{f}^{0}a\right\vert \ll1$ the
non-renormalized and renormalized values are in good agreement, but as
$k_{f}^{0}a\rightarrow-1.21$ the Fermi-pseudopotential diverges away from the
renormalized interaction results. In fact, Ref. \cite{rittenhouse2006hdd}
predicts a collapse of the gas for non-renormalized interactions at $k_{f}%
^{0}a=-1.21$. Just before the point of collapse the ground state energy is
predicted to be $E=\sqrt{5}E_{NI}/3=0.745E_{NI}$. Not only does the
renormalized effective potential cut off the collapse behavior, it also allows
the gas to reach a lower energy than it would be able to without the density
dependence. In other words, if the interaction coefficient in $U_{int}\left(
\vec{r}_{ij}\right)  $ were not density-dependent, but merely involved a cut
off as $k_{f}^{0}a\rightarrow-\infty,$ the gas would not be able to reach the
unitarity energy before collapsing.

\subsection{Unitarity regime}

In this section we explore the behavior of $V_{eff}$ in the strong interaction
regime, i.e. $a\rightarrow\pm\infty$. Examining Eq. \ref{Vintfinal} shows that
the unitarity limit is when $\left\vert k_{f}^{0}a\right\vert \gg R^{\prime}%
$.
\begin{align*}
\dfrac{\left\langle \Phi_{\lambda}\left\vert V_{int}\left(  R_{0}^{\prime
},\Omega\right)  \right\vert \Phi_{\lambda}\right\rangle }{E_{NI}}  &
\rightarrow\dfrac{256\zeta_{\pm}}{9\pi^{2}R^{\prime2}}\int_{0}^{1}y^{6}%
\sqrt{1-y^{2}}dy\\
&  =\dfrac{5\zeta_{\pm}}{9\pi R^{\prime2}}%
\end{align*}
where $\zeta_{\pm}$ is the maximum ($+$) or minimum ($-$) value acquired by
the interaction function $\zeta\left(  k_{f}a\right)  $. This gives a total
effective potential of%
\begin{align}
\dfrac{V_{eff}\left(  R^{\prime}\right)  }{E_{NI}}  &  \rightarrow\dfrac{1}%
{2}R^{\prime2}+\dfrac{1/2+5\zeta_{\pm}/9\pi}{R^{\prime2}}\label{VeffUnit}\\
&  =\left\{
\begin{array}
[c]{ccc}%
\dfrac{1}{2}R^{\prime2}+\dfrac{0.886}{R^{\prime2}} & \text{for} & k_{f}%
^{0}a\gg R^{\prime}\\
&  & \\
\dfrac{1}{2}R^{\prime2}+\dfrac{0.254}{R^{\prime2}} & \text{for} & -k_{f}%
^{0}a\gg R^{\prime}%
\end{array}
\right.  .\nonumber
\end{align}
The hyperradius is a \emph{collective} coordinate so that, as $R\rightarrow0$,
all of the atoms in the system are forced to the center of the trap, which
increases the density of the system. Thus, for small hyperradii, we expect
$V_{eff}$ to act like Eq. \ref{VeffUnit}. In fact, the dashed curves in Figs.
\ref{Veffrepul} and \ref{Veffattract} show that as $R^{\prime}\rightarrow0$
the renormalized effective potential curves start to behave the same,
independently of $a$. Alternatively if the two-body scattering length
approaches $-\infty$, e.g. near a resonance, then we can expect the
interaction to approach Eq. \ref{VeffUnit} for all hyperradii.

In the case where $k_{f}^{0}a\rightarrow\pm\infty$ the effective potential
takes on the form of Eq. \ref{VeffUnit}. Minimization of $V_{eff}$ as a
function of $R^{\prime}$ gives a ground state energy:%
\begin{align}
\dfrac{E}{E_{NI}}  &  =\sqrt{1+10\zeta_{\pm}/9\pi}\label{Eunit}\\
&  =\left\{
\begin{array}
[c]{ccc}%
1.331 & for & a\rightarrow\infty\\
0.712 & for & a\rightarrow-\infty
\end{array}
\right.  .\nonumber
\end{align}
The average hyperradius of the gas is described by the value of the
hyperradius at this minimum which is given by%
\begin{align}
R_{\min}^{\prime}  &  =\left(  1+10\zeta_{\pm}/9\pi\right)  ^{1/4}%
\label{Runit}\\
&  =\left\{
\begin{array}
[c]{ccc}%
1.154 & for & a\rightarrow\infty\\
0.844 & for & a\rightarrow-\infty
\end{array}
\right. \nonumber
\end{align}
At first glance this may seem strange, one might expect the behavior to be
smooth across a resonance, and the energy to connect smoothly from the
$a\rightarrow-\infty$ limit to the $a\rightarrow\infty$ limit
\cite{astrakharchik2004esf}. But the density-dependent renormalization used
here only applies to a degenerate Fermi gas of atoms and does not allow for
the incorporation of higher order correlations, i.e. the formation of diatomic
molecules. Presumably there is another branch in the renormalization that will
match continuously with the $a\rightarrow-\infty$ limit (for a more complete
discussion see section II of Ref. \cite{vonstecher2006renorm}).

Another quantity of interest is the chemical potential of the interacting gas
at unitarity, given by
\[
\mu_{u}=\dfrac{\hbar^{2}k_{f}^{2}\left(  0\right)  }{2m}\left(  1+\beta
\right)  ,
\]
where $\beta$ is a universal parameter. From the single spin component density
given in Eq. \ref{dense} we find that the interacting peak Fermi wavenumber is%
\begin{equation}
k_{f}\left(  0\right)  =\dfrac{k_{f}^{0}}{R_{\min}^{\prime2}}. \label{unitwn}%
\end{equation}
Further, from Eqs. \ref{Eunit} and \ref{Runit}, the ratio of the chemical
potential of the interacting unitarity-limit gas to that of the
non-interacting gas can be written in terms of the rescaled hyperradius as:%
\begin{equation}
\dfrac{\mu_{u}}{\mu}=\dfrac{E}{E_{NI}}=R_{\min}^{\prime2}. \label{unitchem}%
\end{equation}
Solution of Eqs. \ref{unitwn} and \ref{unitchem} in the $a\rightarrow-\infty$
limit yields%
\[
\beta=10\zeta_{-}/9\pi=-0.49.
\]
This value of $\beta$ coincides, not surprisingly, with the value predicted by
the renormalized Hartree-Fock calculation of Ref. \cite{vonstecher2006renorm}.
Even though neither Ref. \cite{vonstecher2006renorm} nor our present treatment
explicitly incorporates Cooper-type fermion pairing, this unitarity limit
$\beta$ is in fair agreement with quantum Monte Carlo estimates that have
obtained $\beta=-0.58$ (Ref. \cite{astrakharchik2004esf}.) and $-0.56$
(Ref.\cite{chang2004qmc} ).

\subsection{Breathing mode excitations}

\begin{figure}[ptb]
\begin{center}
\includegraphics[width=3in]{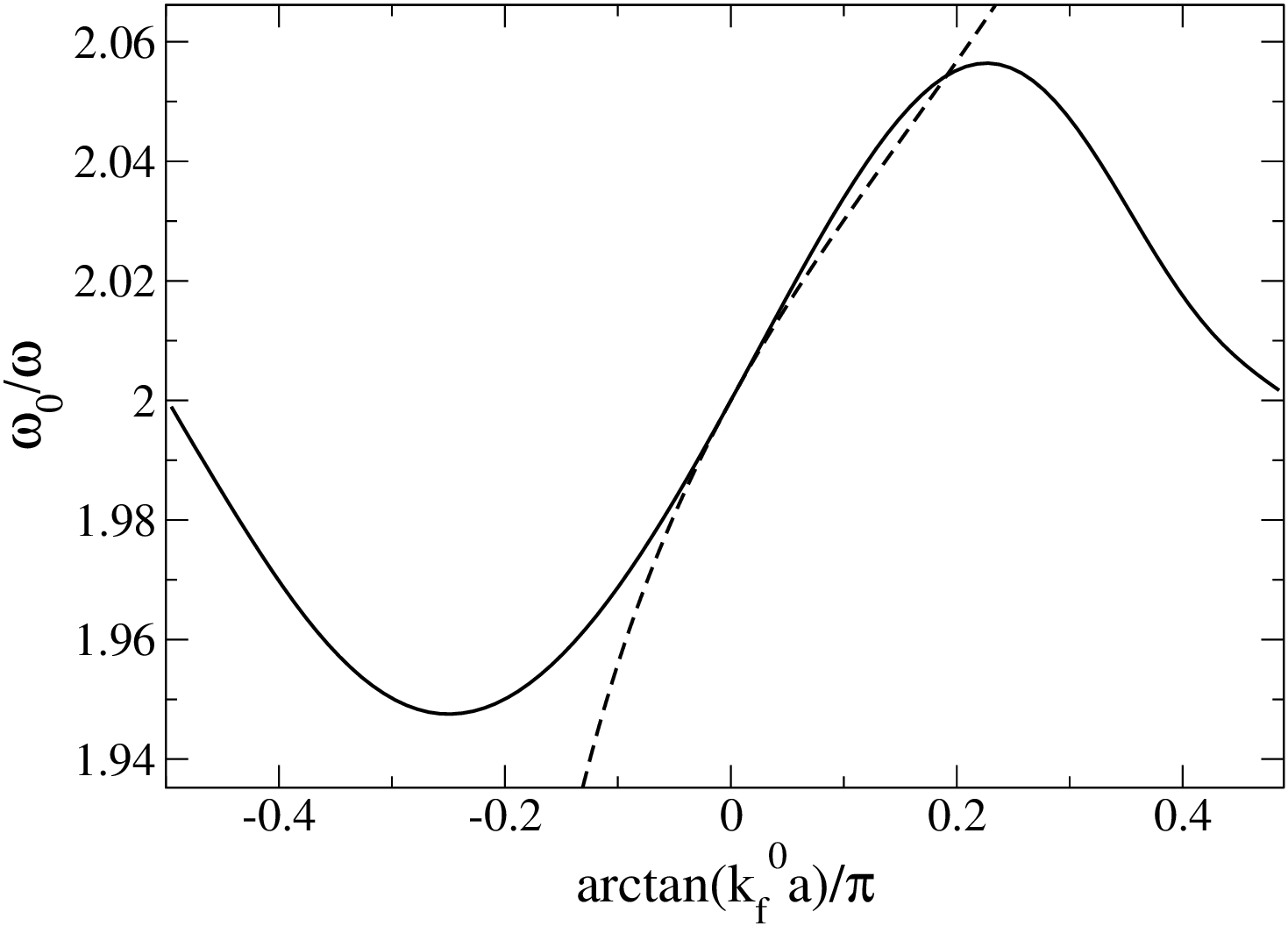}
\end{center}
\caption{The breathing mode frequency $\omega_{0}$ is shown in units of the
trap frequency $\omega$ versus $\arctan\left(  k_{f}^{0}a\right)  /\pi$ . The
solid curve shows the breathing mode predicted using the renormalized
interaction while the dashed curve shows the prediction based on the bare
Fermi pseudo-potential.}%
\label{breath}%
\end{figure}

Refs. \cite{bohn_esry_greene_hsbec,rittenhouse2006hdd} showed that one of the
strengths of the K harmonic method is its ability to predict the lowest radial
excitation, the breathing mode. With the effective potential derived here,
this frequency is found by simply examining the second order Taylor series
about the minimum in $V_{eff}\left(  R^{\prime}\right)  $ and comparing the
resulting Hamiltonian to that of an oscillator. The approximate Hamiltonian is
given by%
\begin{equation}
H=\dfrac{-1}{2m^{\ast}}\dfrac{d^{2}}{dR^{\prime2}}+\dfrac{E_{GS}}{E_{NI}%
}+\dfrac{1}{2E_{NI}}\left.  \dfrac{\partial^{2}V_{eff}}{\partial R^{\prime2}%
}\right\vert _{R^{\prime}=R_{\min}^{\prime}}\left(  R^{\prime}-R_{\min
}^{\prime}\right)  ^{2} \label{Htayor}%
\end{equation}
where $E_{GS}$ is the ground state energy of the system and $R_{\min}^{\prime
}$ is the hyperradius that minimizes $V_{eff}$ scaled by $\sqrt{\left\langle
R^{2}\right\rangle _{NI}}$. this can be recast as the oscillator Hamiltonian%
\begin{equation}
H_{ho}=\dfrac{-1}{2m^{\ast}}\dfrac{d^{2}}{dR^{\prime2}}+\dfrac{E_{GS}}{E_{NI}%
}+\dfrac{1}{2}m^{\ast}\omega_{0}^{\prime2}\left(  R^{\prime}-R_{\min}^{\prime
}\right)  ^{2}. \label{Hho}%
\end{equation}
Here $\omega_{0}^{\prime}$ is given by the second derivative of the $V_{eff}$
at the minimum:%
\begin{subequations}
\begin{equation}
\omega_{0}^{\prime}=\sqrt{\dfrac{1}{m^{\ast}E_{NI}}\left.  \dfrac{\partial
^{2}V_{eff}}{\partial R^{\prime2}}\right\vert _{R^{\prime}=R_{\min}^{\prime}}%
}. \label{BM1}%
\end{equation}
This parameter is the breathing mode frequency in units of $E_{NI}/\hbar$.
Taking $m^{\ast}\rightarrow E_{NI}^{2}/\hbar^{2}\omega^{2}$ in the large $N$
limit gives the breathing mode frequency, $\omega_{0}$:%
\end{subequations}
\begin{equation}
\omega_{0}=\omega\sqrt{\dfrac{1}{E_{NI}}\left.  \dfrac{\partial^{2}V_{eff}%
}{\partial R^{\prime2}}\right\vert _{R^{\prime}=R_{\min}^{\prime}}}
\label{BM2}%
\end{equation}
where $\omega$ is the oscillator frequency of the trap. Fig. \ref{breath}
shows the breathing mode frequency in units of the oscillator frequency
compared to that calculated using non-renormalized interactions. Of course the
breathing mode frequency in the non-interacting limit is $\omega_{0}=2\omega$,
but surprisingly, the frequency turns over and returns to the non-interacting
value as $k_{f}^{0}a\rightarrow\pm\infty$. Upon inserting the second
derivative of Eq. \ref{VeffUnit}, the effective potential as $k_{f}%
^{0}a\rightarrow\pm\infty$ we obtain
\begin{equation}
\omega_{0}=\omega\sqrt{\dfrac{1}{E_{NI}}\left(  1+\left.  3\dfrac
{1+10\zeta_{\pm}/9\pi}{R^{\prime4}}\right\vert _{R^{\prime}=R_{\min}^{\prime}%
}\right)  }. \label{BMunit}%
\end{equation}
When the minimum hyperradius from Eq. \ref{Runit} is plugged in, this implies
that the unitarity limits for the breathing mode frequency are both
$\omega_{0}=2\omega$. This unitarity behavior has also been predicted in Ref.
\cite{werner2006ugi}.

\section{Multiple spin components}

Next consider what happens when the atoms in the gas are equally distributed
among an arbitrary number $\chi$ of spin substates. First, we assume, in order
to limit parameter space, that the s-wave scattering length between two atoms
in any two different spin states has the same value, $a$. Also we neglect the
possibility of inelastic collisions, e.g. of the type:%
\[
\left\vert m_{1}=\dfrac{3}{2}\right\rangle +\left\vert m_{2}=-\dfrac{3}%
{2}\right\rangle \rightarrow\left\vert m_{1}^{\prime}=\dfrac{1}{2}%
\right\rangle +\left\vert m_{2}^{\prime}=-\dfrac{1}{2}\right\rangle .
\]
To proceed, we must also address the question of what density should go into
the renormalized interactions. A particle in spin state $i$ cannot interact
with any other particle in the same spin state by the zero range
approximation, but the density that determines $k_{f}\left(  r\right)  $ in
the renormalization function $\zeta$ might be chosen in various alternative
ways, and we have not yet developed a unique criterion to specify the
appropriate renormalization in this context. As an initial exploration, we
make the assumption that the density of the component that particle $i$ is
interacting with is the density that modifies the interaction, i.e.%
\[
V\left(  r_{ij}\right)  =\dfrac{4\pi\hbar^{2}\zeta\left[  k_{f}^{\left(
j\right)  }\left(  \vec{r}_{i}\right)  a\right]  }{k_{f}^{\left(  j\right)
}\left(  \vec{r}_{i}\right)  }\delta\left(  \vec{r}_{i}-\vec{r}_{j}\right)
\]
where $k_{f}^{\left(  j\right)  }\left(  \vec{r}\right)  =\left[  6\pi^{2}%
\rho^{\left(  j\right)  }\left(  \vec{r}\right)  \right]  ^{1/3}$ is the Fermi
wave number of the spin component that particle $j$ belongs to.

The derivation following this assumptions the same as that for the
two-component gas, up to Eq. \ref{Vint1}. The only added pieces of information
needed are the common density of each component in the effective trap with
oscillator length $l_{eff}$, the chemical potential, the non-interacting
ground state energy and the average hyperradius squared for the system with
$\chi$ spin substates in the large $N$ limit:%
\begin{align}
\rho_{l_{eff}}^{\left(  1\right)  }\left(  \vec{r}\right)   &  =\dfrac{1}%
{6\pi^{2}l_{eff}^{3}}\left(  2\mu\right)  ^{3/2}\left(  1-\dfrac{r^{2}%
}{2l_{eff}^{2}\mu}\right)  ^{3/2}\label{RNImult}\\
\mu &  =\left(  \dfrac{6N}{\chi}\right)  ^{1/3}\\
E_{NI}  &  =\hbar\omega\dfrac{\left(  6N\right)  ^{4/3}}{\chi^{1/3}%
8}\label{ENI}\\
\left\langle R^{2}\right\rangle _{NI}  &  =\dfrac{\hbar}{m\omega}%
\dfrac{\left(  6N\right)  ^{4/3}}{\chi^{1/3}8N}.
\end{align}
The sum over spin substates in Eq. \ref{Vint1} results in a factor of
$\chi\left(  \chi-1\right)  $, leaving%
\begin{equation}
\left\langle \Phi_{\lambda}\left\vert V_{int}\left(  R^{\prime},\Omega\right)
\right\vert \Phi_{\lambda}\right\rangle =\dfrac{4\pi\hbar^{2}}{m}\dfrac
{\chi\left(  \chi-1\right)  }{2}\int\dfrac{\zeta\left[  k_{f}\left(  \vec
{r}\right)  a\right]  }{k_{f}\left(  \vec{r}\right)  }\left[  \rho_{l_{eff}%
}^{\left(  1\right)  }\left(  \vec{r}\right)  \right]  ^{2}d^{3}r.
\label{Vint2}%
\end{equation}
Another change of integration variables which gives the effective hyperradial
potential as%
\begin{align}
\dfrac{V_{eff}\left(  R^{\prime}\right)  }{E_{NI}}  &  =\dfrac{1}{2R^{\prime
2}}+\dfrac{1}{2}R^{\prime2}+\left(  \chi-1\right)  \dfrac{64N^{4/3}}%
{3^{2/3}\pi^{2}R^{\prime2}}f\left(  \dfrac{k_{f}^{0}a}{R^{\prime}}\right)
.\label{Veffmulti}\\
f\left(  \dfrac{k_{f}^{0}a}{R^{\prime}}\right)   &  \equiv\int_{0}^{1}%
y^{6}\sqrt{1-y^{2}}\zeta\left(  \dfrac{k_{f}^{0}a}{R^{\prime}}y\right)
dy.\nonumber
\end{align}
Comparison with Eq. \ref{Vefffinal}, the effective potential for the two
component gas, demonstrates that the extra spin components increase the
strength of the interaction by a factor of $\chi-1$.

In the limit where $k_{f}^{0}a\rightarrow-\infty$ the effective potential
limits to%
\begin{equation}
\dfrac{V_{eff}\left(  R^{\prime}\right)  }{E_{NI}}\rightarrow\dfrac{1}%
{2}R^{\prime2}+\dfrac{1+10\left(  \chi-1\right)  \zeta_{-}/9\pi}{2R^{\prime2}}
\label{Veffmultiunit}%
\end{equation}
Taking $\chi=3$ yields%
\begin{align}
\dfrac{V_{eff}\left(  R^{\prime}\right)  }{E_{NI}}  &  \rightarrow\dfrac{1}%
{2}R^{\prime2}+\dfrac{1-20\zeta_{-}/9\pi}{2R^{\prime2}}\label{Veff3unit}\\
&  =\dfrac{1}{2}R^{\prime2}+\dfrac{0.00772}{R^{\prime2}}.\nonumber
\end{align}
In this limit the barrier preventing the gas from falling in to the center of
the trap, i.e. $R^{\prime}\rightarrow0$, is \emph{very }weak. The unitarity
energy and average hyperradius are given by%
\begin{align*}
\allowbreak\dfrac{E}{E_{NI}}  &  =\sqrt{1+20\zeta_{-}/9\pi}=0.124\\
\dfrac{R_{\min}}{\sqrt{\left\langle R^{2}\right\rangle _{NI}}}  &  =\left(
1+20\zeta_{-}/9\pi\right)  ^{1/4}=0.352.
\end{align*}
Since the K harmonic method is intrinsically a variational calculation, it is
very possible that a better calculation, for example using the Hartree-Fock
method, might show that the 3 component gas becomes mechanically unstable in
the unitarity limit \cite{heiselberg2001fsl}. In other words the three
component gas might collapse in a manner similar to that of the Bosenova
\cite{bohn_esry_greene_hsbec,shuryak1996mbc,stoof1997mqt,bradley1995ebe,donley2001dca}%
.

With $\chi=4$ the effective potential at $k_{f}^{0}a\rightarrow-\infty$
becomes entirely attractive%
\begin{align}
\dfrac{V_{eff}\left(  R^{\prime}\right)  }{E_{NI}} &  \rightarrow\dfrac{1}%
{2}R^{\prime2}+\dfrac{1-30\zeta_{-}/9\pi}{2R^{\prime2}}\label{Veff4unit}\\
&  =\dfrac{1}{2}R^{\prime2}-\dfrac{\allowbreak0.238}{R^{\prime2}}\nonumber
\end{align}
meaning that the gas is predicted to collapse down toward $R^{\prime
}\rightarrow0$. Presumably some very rich and complex dynamics (cluster
formation, inelastic collisions, etc.) occur during this process, but our K
harmonic trial wave function is too simple to describe these phenomena. When
the local minimum in $V_{eff}\left(  R^{\prime}\right)  $ becomes a saddle
point the gas is no longer mechanically stable and is free to collapse. This
occurs at a critical interaction strength of $k_{f}a_{c}=-0.657$.

\section{Summary and prospects}

We have shown that applying the variational hyperspherical treatment of Ref.
\cite{rittenhouse2006hdd} to a density-dependent, zero-range, s-wave
interaction produces a unique, physically intuitive picture of the behavior of
gas. By fixing the hyperangular behavior of the gas, in the large atom number
limit, a simple 1D effective potential in a collective coordinate, the
hyperradius $R$ was produced. The ground state energy and rms radius of the
two component gas predicted by this method are in excellent agreement with
those predicted using the Hartree-Fock method. The ground state energy of the
two component gas goes to a finite, unitarity limit as the two-body scattering
length $a$ diverges to $+\infty$. Due to the inability of our hyperspherical
trial function and the renormalized, density-dependent interaction to describe
two-body bound states, the method only applies (at this level of our
development) to a degenerate Fermi gas of atoms. In the $a\rightarrow-\infty$
the two component gas has a stable ground state and avoids the collapsing
behavior predicted for the bare Fermi pseudo-potential, and has a ground state
energy of $0.73$ times the ground state energy of $N$ non-interacting fermions
in an isotropic oscillator trap. This energy is lower than the minimum energy
that could be reached before collapse, which would be predicted without
density-dependent interactions. In the unitarity limit, the lowest radial
excitation frequency, the breathing mode frequency, for the two component gas
is the \emph{same} as the non-interacting value in both the positive and
negative asymptotic scattering length limits.

The effective potential for a three component gas has a weak hyperradial
barrier in the $a\rightarrow-\infty$ limit, which just barely prevents
collapse of the gas. Because of the variational nature of the K-harmonic
method, a better approximation of the wavefunction, as in the Hartree-Fock
method, might conceivably allow the three component system to collapse. For
four or more spin components the repulsive barrier becomes attractive in the
$a\rightarrow-\infty$ and the gas is predicted to collapse in a manner similar
to the Bosenova
\cite{bohn_esry_greene_hsbec,shuryak1996mbc,stoof1997mqt,bradley1995ebe,donley2001dca}%
. Some uncertainty about this prediction still exists, however, because we
have not yet validated our renormalization procedure for DFGs containing more
than two spin components. Here, the interactions between different spin
components was assumed to be the same for all possible combinations. A more
accurate treatment would include different two-body scattering lengths for
different combinations of spin components, though if all scattering lengths
are large and negative, the prediction of instability would still apply.
Higher-order correlations, such as BEC or BCS pairing are beyond the scope of
this work, and relegated to future publication.

\section*{ACKNOWLEDGMENTS}

We are indebted to Javier von Stecher for extensive discussions, and for
providing some of the Hartree-Fock results that are shown in this paper for
comparison. One of us (CHG) received partial support from the Miller Institute
for Basic Research in Science, University of California Berkeley. This work
was also supported in part by funding from the NSF.


\begin{thebibliography}{26}
\expandafter\ifx\csname natexlab\endcsname\relax\def\natexlab#1{#1}\fi
\expandafter\ifx\csname bibnamefont\endcsname\relax
  \def\bibnamefont#1{#1}\fi
\expandafter\ifx\csname bibfnamefont\endcsname\relax
  \def\bibfnamefont#1{#1}\fi
\expandafter\ifx\csname citenamefont\endcsname\relax
  \def\citenamefont#1{#1}\fi
\expandafter\ifx\csname url\endcsname\relax
  \def\url#1{\texttt{#1}}\fi
\expandafter\ifx\csname urlprefix\endcsname\relax\def\urlprefix{URL }\fi
\providecommand{\bibinfo}[2]{#2}
\providecommand{\eprint}[2][]{\url{#2}}

\bibitem[{\citenamefont{Fermi}(1936)}]{fermi1936}
\bibinfo{author}{\bibfnamefont{E.}~\bibnamefont{Fermi}}, \bibinfo{journal}{Ric.
  Sci.} \textbf{\bibinfo{volume}{7}}, \bibinfo{pages}{13}
  (\bibinfo{year}{1936}).

\bibitem[{\citenamefont{Bruun and Burnett}(1998)}]{bruun1998ifg}
\bibinfo{author}{\bibfnamefont{G.M.}~\bibnamefont{Bruun}} \bibnamefont{and}
  \bibinfo{author}{\bibfnamefont{K.}~\bibnamefont{Burnett}},
  \bibinfo{journal}{Phys. Rev. A} \textbf{\bibinfo{volume}{58}},
  \bibinfo{pages}{2427} (\bibinfo{year}{1998}).

\bibitem[{\citenamefont{Houbiers et~al.}(1997)\citenamefont{Houbiers, Ferwerda,
  Stoof, McAlexander, Sackett, and Hulet}}]{houbiers1997ssa}
\bibinfo{author}{\bibfnamefont{M.}~\bibnamefont{Houbiers}},
  \bibinfo{author}{\bibfnamefont{R.}~\bibnamefont{Ferwerda}},
  \bibinfo{author}{\bibfnamefont{H.T.C.}~\bibnamefont{Stoof}},
  \bibinfo{author}{\bibfnamefont{W.I.}~\bibnamefont{McAlexander}},
  \bibinfo{author}{\bibfnamefont{C.A.}~\bibnamefont{Sackett}}, \bibnamefont{and}
  \bibinfo{author}{\bibfnamefont{R.G.}~\bibnamefont{Hulet}},
  \bibinfo{journal}{Phys. Rev. A} \textbf{\bibinfo{volume}{56}},
  \bibinfo{pages}{4864} (\bibinfo{year}{1997}).

\bibitem[{\citenamefont{Roth and Feldmeier}(2001)}]{roth2001esa}
\bibinfo{author}{\bibfnamefont{R.}~\bibnamefont{Roth}} \bibnamefont{and}
  \bibinfo{author}{\bibfnamefont{H.}~\bibnamefont{Feldmeier}},
  \bibinfo{journal}{Phys. Rev. A} \textbf{\bibinfo{volume}{64}},
  \bibinfo{pages}{43603} (\bibinfo{year}{2001}).

\bibitem[{\citenamefont{Bartenstein et~al.}(2004)\citenamefont{Bartenstein,
  Altmeyer, Riedl, Jochim, Chin, Denschlag, and Grimm}}]{bartenstein2004cmb}
\bibinfo{author}{\bibfnamefont{M.}~\bibnamefont{Bartenstein}},
  \bibinfo{author}{\bibfnamefont{A.}~\bibnamefont{Altmeyer}},
  \bibinfo{author}{\bibfnamefont{S.}~\bibnamefont{Riedl}},
  \bibinfo{author}{\bibfnamefont{S.}~\bibnamefont{Jochim}},
  \bibinfo{author}{\bibfnamefont{C.}~\bibnamefont{Chin}},
  \bibinfo{author}{\bibfnamefont{J.H.}~\bibnamefont{Denschlag}},
  \bibnamefont{and} \bibinfo{author}{\bibfnamefont{R.}~\bibnamefont{Grimm}},
  \bibinfo{journal}{Phys. Rev. Lett.} \textbf{\bibinfo{volume}{92}},
  \bibinfo{pages}{120401} (\bibinfo{year}{2004}).

\bibitem[{\citenamefont{Bourdel et~al.}(2004)\citenamefont{Bourdel, Khaykovich,
  Cubizolles, Zhang, Chevy, Teichmann, Tarruell, Kokkelmans, and
  Salomon}}]{bourdel2004esb}
\bibinfo{author}{\bibfnamefont{T.}~\bibnamefont{Bourdel}},
  \bibinfo{author}{\bibfnamefont{L.}~\bibnamefont{Khaykovich}},
  \bibinfo{author}{\bibfnamefont{J.}~\bibnamefont{Cubizolles}},
  \bibinfo{author}{\bibfnamefont{J.}~\bibnamefont{Zhang}},
  \bibinfo{author}{\bibfnamefont{F.}~\bibnamefont{Chevy}},
  \bibinfo{author}{\bibfnamefont{M.}~\bibnamefont{Teichmann}},
  \bibinfo{author}{\bibfnamefont{L.}~\bibnamefont{Tarruell}},
  \bibinfo{author}{\bibfnamefont{S.J.J.M.F.}~\bibnamefont{Kokkelmans}},
  \bibnamefont{and} \bibinfo{author}{\bibfnamefont{C.}~\bibnamefont{Salomon}},
  \bibinfo{journal}{Phys. Rev. Lett.} \textbf{\bibinfo{volume}{93}},
  \bibinfo{pages}{50401} (\bibinfo{year}{2004}).

\bibitem[{\citenamefont{Kinast et~al.}(2004)\citenamefont{Kinast, Hemmer, Gehm,
  Turlapov, and Thomas}}]{kinast2004esr}
\bibinfo{author}{\bibfnamefont{J.}~\bibnamefont{Kinast}},
  \bibinfo{author}{\bibfnamefont{S.L.}~\bibnamefont{Hemmer}},
  \bibinfo{author}{\bibfnamefont{M.E.}~\bibnamefont{Gehm}},
  \bibinfo{author}{\bibfnamefont{A.}~\bibnamefont{Turlapov}}, \bibnamefont{and}
  \bibinfo{author}{\bibfnamefont{J.E.}~\bibnamefont{Thomas}},
  \bibinfo{journal}{Phys. Rev. Lett.} \textbf{\bibinfo{volume}{92}},
  \bibinfo{pages}{150402} (\bibinfo{year}{2004}).

\bibitem[{\citenamefont{Regal et~al.}(2004)\citenamefont{Regal, Greiner, and
  Jin}}]{regal2004orc}
\bibinfo{author}{\bibfnamefont{C.A.}~\bibnamefont{Regal}},
  \bibinfo{author}{\bibfnamefont{M.}~\bibnamefont{Greiner}}, \bibnamefont{and}
  \bibinfo{author}{\bibfnamefont{D.S.}~\bibnamefont{Jin}},
  \bibinfo{journal}{Phys. Rev. Lett.} \textbf{\bibinfo{volume}{92}},
  \bibinfo{pages}{40403} (\bibinfo{year}{2004}).

\bibitem[{\citenamefont{Zwierlein et~al.}(2004)\citenamefont{Zwierlein, Stan,
  Schunck, Raupach, Kerman, and Ketterle}}]{zwierlein2004cpf}
\bibinfo{author}{\bibfnamefont{M.W.}~\bibnamefont{Zwierlein}},
  \bibinfo{author}{\bibfnamefont{C.A.}~\bibnamefont{Stan}},
  \bibinfo{author}{\bibfnamefont{C.H.}~\bibnamefont{Schunck}},
  \bibinfo{author}{\bibfnamefont{S.M.F.}~\bibnamefont{Raupach}},
  \bibinfo{author}{\bibfnamefont{A.J.}~\bibnamefont{Kerman}}, \bibnamefont{and}
  \bibinfo{author}{\bibfnamefont{W.}~\bibnamefont{Ketterle}},
  \bibinfo{journal}{Phys. Rev. Lett.} \textbf{\bibinfo{volume}{92}},
  \bibinfo{pages}{120403} (\bibinfo{year}{2004}).

\bibitem[{\citenamefont{von Stecher and Greene}(2006)}]{vonstecher2006renorm}
\bibinfo{author}{\bibfnamefont{J.}~\bibnamefont{von Stecher}} \bibnamefont{and}
  \bibinfo{author}{\bibfnamefont{C.~H.} \bibnamefont{Greene}},
  \bibinfo{journal}{eprint cond-mat/0610848}  (\bibinfo{year}{2006}).

\bibitem[{\citenamefont{Rittenhouse et~al.}(2006)\citenamefont{Rittenhouse,
  Cavagnero, von Stecher, and Greene}}]{rittenhouse2006hdd}
\bibinfo{author}{\bibfnamefont{S.T.}~\bibnamefont{Rittenhouse}},
  \bibinfo{author}{\bibfnamefont{M.J.}~\bibnamefont{Cavagnero}},
  \bibinfo{author}{\bibfnamefont{J.}~\bibnamefont{von Stecher}},
  \bibnamefont{and} \bibinfo{author}{\bibfnamefont{C.H.}~\bibnamefont{Greene}},
  \bibinfo{journal}{Phys. Rev. A} \textbf{\bibinfo{volume}{74}},
  \bibinfo{pages}{053624} (\bibinfo{year}{2006}).

\bibitem[{\citenamefont{Avery}(1989)}]{avery}
\bibinfo{author}{\bibfnamefont{J.}~\bibnamefont{Avery}},
  \emph{\bibinfo{title}{{Hyperspherical Harmonics: Applications in Quantum
  Theory}}} (\bibinfo{publisher}{Kluwer Academic Publishers},
  \bibinfo{address}{Norwell, MA}, \bibinfo{year}{1989}).

\bibitem[{\citenamefont{Fano and Rau}(1986)}]{fano1986aca}
\bibinfo{author}{\bibfnamefont{U.}~\bibnamefont{Fano}} \bibnamefont{and}
  \bibinfo{author}{\bibfnamefont{A.}~\bibnamefont{Rau}},
  \emph{\bibinfo{title}{Atomic Collisions and Spectra}}
  (\bibinfo{publisher}{Academic Press}, \bibinfo{address}{Orlando, FL},
  \bibinfo{year}{1986}).

\bibitem[{\citenamefont{Smirnov and Shitikova}(1977)}]{SmirnovShitikova}
\bibinfo{author}{\bibfnamefont{Y.~F.} \bibnamefont{Smirnov}} \bibnamefont{and}
  \bibinfo{author}{\bibfnamefont{K.~V.} \bibnamefont{Shitikova}},
  \bibinfo{journal}{Sov. J. Part. Nucl.} \textbf{\bibinfo{volume}{8}},
  \bibinfo{pages}{44} (\bibinfo{year}{1977}).

\bibitem[{\citenamefont{Bohn et~al.}(1998)\citenamefont{Bohn, Esry, and
  Greene}}]{bohn_esry_greene_hsbec}
\bibinfo{author}{\bibfnamefont{J.~L.} \bibnamefont{Bohn}},
  \bibinfo{author}{\bibfnamefont{B.~D.} \bibnamefont{Esry}}, \bibnamefont{and}
  \bibinfo{author}{\bibfnamefont{C.~H.} \bibnamefont{Greene}},
  \bibinfo{journal}{Phys. Rev. A} \textbf{\bibinfo{volume}{58}},
  \bibinfo{pages}{584} (\bibinfo{year}{1998}).

\bibitem[{\citenamefont{Timofeyuk}(2004)}]{Timofeyuk04}
\bibinfo{author}{\bibfnamefont{N.~K.} \bibnamefont{Timofeyuk}},
  \bibinfo{journal}{Phys. Rev. C} \textbf{\bibinfo{volume}{69}},
  \bibinfo{pages}{034336} (\bibinfo{year}{2004}).

\bibitem[{dim()}]{dimensionalPT}
\bibinfo{note}{See, for instance D. Z. Goodson, M. Lopez-Cabrera, D. R.
  Herschbach, J. D. Morgan III, J. Chem. Phys.{ \bf {97}}, 8481 (1992); J. G.
  Loeser, J. H. Summerfield, A. L. Tan, Z. Zheng, J. Chem. Phys. {\bf{100}},
  5036 (1994); M. Dunn and D. K. Watson, Annals of Physics {\bf {251}}, 266-318
  and 319-336 (1996).}

\bibitem[{\citenamefont{Cowan}(1981)}]{cowan1981tas}
\bibinfo{author}{\bibfnamefont{R.~D.} \bibnamefont{Cowan}},
  \emph{\bibinfo{title}{The Theory of Atomic Structure and Spectra}}
  (\bibinfo{publisher}{University of California Press}, \bibinfo{address}{Los
  Angeles, CA}, \bibinfo{year}{1981}).

\bibitem[{\citenamefont{Astrakharchik et~al.}(2004)\citenamefont{Astrakharchik,
  Boronat, Casulleras, and Giorgini}}]{astrakharchik2004esf}
\bibinfo{author}{\bibfnamefont{G.E.}~\bibnamefont{Astrakharchik}},
  \bibinfo{author}{\bibfnamefont{J.}~\bibnamefont{Boronat}},
  \bibinfo{author}{\bibfnamefont{J.}~\bibnamefont{Casulleras}},
  \bibnamefont{and} \bibinfo{author}{\bibfnamefont{S.}~\bibnamefont{Giorgini}},
  \bibinfo{journal}{Phys. Rev. Lett.} \textbf{\bibinfo{volume}{93}},
  \bibinfo{pages}{200404} (\bibinfo{year}{2004}).

\bibitem[{\citenamefont{Chang et~al.}(2004)\citenamefont{Chang, Pandharipande,
  Carlson, and Schmidt}}]{chang2004qmc}
\bibinfo{author}{\bibfnamefont{S.Y}~\bibnamefont{Chang}},
  \bibinfo{author}{\bibfnamefont{V.R.}~\bibnamefont{Pandharipande}},
  \bibinfo{author}{\bibfnamefont{J.}~\bibnamefont{Carlson}}, \bibnamefont{and}
  \bibinfo{author}{\bibfnamefont{K.E.}~\bibnamefont{Schmidt}},
  \bibinfo{journal}{Phys. Rev. A} \textbf{\bibinfo{volume}{70}},
  \bibinfo{pages}{43602} (\bibinfo{year}{2004}).

\bibitem[{\citenamefont{Werner and Castin}(2006)}]{werner2006ugi}
\bibinfo{author}{\bibfnamefont{F.}~\bibnamefont{Werner}} \bibnamefont{and}
  \bibinfo{author}{\bibfnamefont{Y.}~\bibnamefont{Castin}},
  \bibinfo{journal}{Phys. Rev. A} \textbf{\bibinfo{volume}{74}},
  \bibinfo{pages}{53604} (\bibinfo{year}{2006}).

\bibitem[{\citenamefont{Heiselberg}(2001)}]{heiselberg2001fsl}
\bibinfo{author}{\bibfnamefont{H.}~\bibnamefont{Heiselberg}},
  \bibinfo{journal}{Phys. Rev. A} \textbf{\bibinfo{volume}{63}},
  \bibinfo{pages}{43606} (\bibinfo{year}{2001}).

\bibitem[{\citenamefont{Bradley et~al.}(1995)\citenamefont{Bradley, Sackett,
  Tollett, and Hulet}}]{bradley1995ebe}
\bibinfo{author}{\bibfnamefont{C.C.}~\bibnamefont{Bradley}},
  \bibinfo{author}{\bibfnamefont{C.A.}~\bibnamefont{Sackett}},
  \bibinfo{author}{\bibfnamefont{J.J.}~\bibnamefont{Tollett}}, \bibnamefont{and}
  \bibinfo{author}{\bibfnamefont{R.G.}~\bibnamefont{Hulet}},
  \bibinfo{journal}{Phys. Rev. Lett.} \textbf{\bibinfo{volume}{75}},
  \bibinfo{pages}{1687} (\bibinfo{year}{1995}).

\bibitem[{\citenamefont{Donley et~al.}(2001)\citenamefont{Donley, Claussen,
  Cornish, Roberts, Cornell, and Wieman}}]{donley2001dca}
\bibinfo{author}{\bibfnamefont{E.}~\bibnamefont{Donley}},
  \bibinfo{author}{\bibfnamefont{N.}~\bibnamefont{Claussen}},
  \bibinfo{author}{\bibfnamefont{S.}~\bibnamefont{Cornish}},
  \bibinfo{author}{\bibfnamefont{J.}~\bibnamefont{Roberts}},
  \bibinfo{author}{\bibfnamefont{E.}~\bibnamefont{Cornell}}, \bibnamefont{and}
  \bibinfo{author}{\bibfnamefont{C.}~\bibnamefont{Wieman}},
  \bibinfo{journal}{Nature} \textbf{\bibinfo{volume}{412}},
  \bibinfo{pages}{295} (\bibinfo{year}{2001}).

\bibitem[{\citenamefont{Shuryak}(1996)}]{shuryak1996mbc}
\bibinfo{author}{\bibfnamefont{E.V.}~\bibnamefont{Shuryak}},
  \bibinfo{journal}{Phys. Rev. A} \textbf{\bibinfo{volume}{54}},
  \bibinfo{pages}{3151} (\bibinfo{year}{1996}).

\bibitem[{\citenamefont{Stoof}(1997)}]{stoof1997mqt}
\bibinfo{author}{\bibfnamefont{H.}~\bibnamefont{Stoof}}, \bibinfo{journal}{J.
  Stat. Phys.} \textbf{\bibinfo{volume}{87}}, \bibinfo{pages}{1353}
  (\bibinfo{year}{1997}).

\end{thebibliography}

\end{document}